\documentclass[a4paper,11pt]{article}
\pdfoutput=1 
\usepackage{jcappub}
\usepackage[normalem]{ulem}
\usepackage[utf8]{inputenc}
\usepackage{enumerate}
\usepackage{amsmath, amssymb, amsthm, graphicx, epsfig, fancyhdr,epsfig, slashed}
\usepackage[mathscr]{euscript}
\let\oldcdot\cdot
\usepackage{breqn}
\let\cdot\oldcdot
\usepackage{epsfig}  
\usepackage{graphicx}   
\usepackage{slashed}       
\usepackage{tikz}
\usetikzlibrary{shapes.geometric, arrows}
\usepackage{subcaption} 
\usepackage{url}
\usepackage{color,soul} 
\usepackage{multirow}
\usepackage{comment}
\usepackage{tabularx}
\usepackage{mathtools}
\usepackage{booktabs}
\usepackage{bm}
\usepackage{colortbl}
\usepackage{hyperref}
\usepackage[all]{hypcap}
\usepackage[verbose]{placeins}
\hypersetup{ 
    colorlinks=true,
    linkcolor=blue,
    filecolor=red,      
    urlcolor=purple,
    citecolor=green,
    }

\title{\bf Exploring neutrino interactions in light of present and upcoming galaxy surveys}
\author[a,1]{Sourav Pal,\note{Corresponding author.}}
\author[b]{Rickmoy Samanta}
\author[a]{and Supratik Pal}

\emailAdd{soupal1729@gmail.com}
\emailAdd{rickmoysamanta@gmail.com}
\emailAdd{supratik@isical.ac.in}

\affiliation[a]{Physics and Applied Mathematics Unit, Indian Statistical Institute,\\ 203, B.T. Road, Kolkata 700108, India}
\affiliation[b]{Department of Physics, Birla Institute of Technology and Science Pilani,\\
Hyderabad 500078, India}

 \abstract
{
In the standard  cosmological framework, neutrinos begin to free-stream after the weak interaction phase ends in the early universe, at a temperature of approximately $T \sim 1 \, {\rm MeV}$. However, the onset of neutrino free-streaming can be delayed if additional interactions occur in the early universe, leaving  imprints on both the cosmic microwave background (CMB) angular power spectra and the large-scale structure (LSS) matter power spectra. We present a thorough analysis of early universe neutrino interactions with a fairly generalized  parameterization of the interaction rates as a power law in neutrino temperature. In this (6+2) parameter scenario, we constrain the cosmological parameters along with the free-streaming redshift and the sum of the neutrino mass in presence of such interactions, with the help of full shape (FS) galaxy power spectra from BOSS Data Release 12. Our analysis reveals that a combined dataset of FS along with CMB and BAO offers improved constraints on the free-streaming redshift from present data, comparable to the forecast results from future CMB-S4 data. Additionally, we investigate the prospects of future galaxy surveys by forecasting on Euclid mission in combination with Planck and CMB-S4, and find significant improvement on both the free-streaming redshift and the sum of the neutrino mass than the existing constraints as well as than CMB-S4 alone.  
}

\begin{document}
\maketitle
\flushbottom

\section{Introduction}
\label{sec: introduction}

Neutrinos, ever-present throughout the universe's cosmic history, hold a profound significance in our understanding of fundamental physics. While they are considered massless in the standard model of particle physics, neutrino oscillation experiments \cite{LSND:1995lje,Super-Kamiokande:1998kpq} conducted over the past few decades have revealed that at least two of the neutrino flavor eigenstates possess mass. Although neutrino oscillation experiments \cite{SNO:2002tuh,IceCubeCollaboration:2023wtb,NOvA:2023uxq} constrain the mass-splitting, but do not yet provide a direct measurements of the absolute neutrino mass. Current and future cosmological observations may chip in here by providing further information about the sum of neutrino mass and possible neutrino interactions. The most stringent upper bound on the sum of neutrino mass currently stands at $\sum m_\nu < 0.072\; {\rm eV}$ \cite{DESI:2024mwx}.
The presence of a non-zero neutrino mass brings forth a wide range of beyond standard model (BSM) scenarios, each allowing for different types of neutrino interactions. Terrestrial experiments have already made remarkable strides in probing these BSM interactions \cite{MiniBooNE:2020pnu,IceCubeCollaboration:2023wtb}. Complimentary to that, cosmological observations provide a wealth of information about these interactions, offering a unique lens through which we can explore both the sum of neutrino mass and the underlying nature of their interactions.

In the standard model of cosmology, neutrinos decouple from the primordial plasma at around $\sim 1 \; \rm{MeV}$, when the weak interaction ($\sim G_{\rm F}^2 \,T^5$) is suppressed in comparison to the Hubble expansion rate $H(T)$. Here $G_{\rm F}$ is the Fermi coupling constant, which denotes the strength of the weak interaction. After decoupling, neutrinos free-stream through the photon-baryon fluid with almost the speed of light, that eventually drags the photon-baryon fluid towards smaller scales. Moreover, neutrinos introduce a significant anisotropic component in the evolution of gravitational potential, leading to a noticeable suppression in the peaks of the Cosmic Microwave Background (CMB) spectrum \cite{Lesgourgues:2006nd,Bashinsky:2003tk,Lesgourgues:2013sjj,Dolgov:2002wy}. These influences extend to the Baryon Acoustic Oscillation (BAO) features observed in the large-scale structure of the late universe \cite{Follin:2015hya,Peloso:2015jua,Baumann:2019keh}. The combined insights from CMB and BAO measurements thus serve to constrain the elusive nature of neutrino free-streaming and corresponding interactions \cite{Baumann:2019keh,Couchot:2017pvz,Tanseri:2022zfe}. 

Various well-motivated neutrino interactions  have been brought forth in the last few decades. These include interactions in both early and late universe. Neutrino interactions in the early universe delay the onset of free-streaming and as a result leaves distinguishable signatures in both the CMB power spectra and BAO. In the early universe, neutrino interactions are preferably described by the four Fermi self-interaction where the interaction strength varies with temperature as $T_\nu^5$. These kinds of neutrino interactions have been studied before in the literature in light of CMB \cite{Cyr-Racine:2013jua,Oldengott:2014qra,Lancaster:2017ksf,Kreisch:2019yzn,Park:2019ibn,RoyChoudhury:2020dmd,Das:2020xke,Lancaster:2017ksf,Choudhury:2021dsc,Das:2021guu,Das:2023npl,Berger:2022cab}. In addition, various BSM interactions within the neutrino sector have also been studied, including scenarios where neutrinos annihilate into massless scalars \cite{Bell:2005dr,Hannestad:2004qu,Venzor:2022hql,Archidiacono:2013dua,Forastieri:2019cuf,Venzor:2023aka} or undergo decay and inverse decay  via eV-scale neutrinophilic scalars \cite{Escudero:2019gvw,Escudero:2019gfk,Chacko:2019nej,Chacko:2020hmh}. These latter interactions are more prominent at lower temperatures and primarily impact large-scale modes, whereas neutrino self-interactions on the other hand are most effective at higher temperatures, influencing small-scale modes.  
Neutrinos interacting at late universe as well as transient interactions have a free-stream window in the redshift range $2000<z<10^5$, primarily affecting the large scale modes \cite{Taule:2022jrz}. On the other hand, small scale modes carry information of interactions that are dominant at high temperature (early universe), which is the focus of the present  analysis.

Most of the previous studies on interacting neutrino models have primarily focused on the CMB. Recently, however, information from the Large Scale Structure (LSS), particularly in the mildly non-linear regime, has been employed to investigate scenarios of strongly interacting (SI) and moderately interacting (MI) modes in $T_\nu^5$ type neutrino self-interactions in \cite{Camarena:2023cku,He:2023oke,Camarena:2024zck}. However, a major challenge in probing the non-linear regime is the breakdown of standard perturbation theory after the liner regime, that calls for simulation-based approaches \cite{Smith:2002dz,bird2012massive,Takahashi:2012em}. Possible alternatives like the Effective Field Theory (EFT) of Large Scale Structure (LSS) has shown promise in extracting cosmological information from small scales, at least up to the mildly non-linear regime. In order to investigate non-trivial neutrino interactions that may not be easily tractable in simulations, our study will follow the EFT of LSS approach. Additionally, previous studies mainly focused on constraining the effective interaction strength in four Fermi like neutrino self-interactions using EFT of LSS. Our analysis adopts a more general parameterization of neutrino interactions in the early universe, following  \cite{Taule:2022jrz}. We adopt a temperature-dependent parameterization of the neutrino interaction rate as detailed in \cite{Taule:2022jrz} and characterize it  by a power law in temperature, $\Gamma_\nu \propto T_\nu^{n_{\rm int}}$, where $\Gamma_\nu$ represents the neutrino interaction rate, $T_\nu$ is the background neutrino temperature, and $n_{\rm int}$ is a power-law index that generalizes all types of neutrino interactions. 
Specifically, we consider interactions with power law index $n_{\rm int} =3,\, 4$ and $5$ respectively in our analysis. This will also help us investigate the scenario in a fairly model-independent way.  Although some  of the scenarios may not be readily mapped into a simple particle physics models, examining the free-streaming window of neutrinos through LSS analysis, in conjunction with recent CMB studies, remains a valuable endeavor. 

In this work, we investigate the effects of neutrino interaction on LSS  following the EFT of LSS approach and search for possible bounds on the parameters from present and upcoming Galaxy Surveys in combination with CMB. Although we constrain the standard cosmological parameters (in the background $\Lambda$CDM setup) along with the neutrino interactions parameters, our primary intention is to find out possible bounds on the interaction redshift and the sum  over neutrino mass. 
Here the term ``interaction redshift'' refers to the specific redshift at which neutrinos start to free-stream  (denoted by $z_{\rm int}$ throughout the paper, which is essentially the decoupling redshift of neutrinos from the specific interaction under consideration.\footnote{ Note that, for interactions active at low redshift regime, neutrinos can recouple again with an unknown scalar field (in eV-scale neutrinophilic model), 
which is not considered here in this analysis.}) 
Additionally, contrary to the previous analysis \cite{Taule:2022jrz}, we make use of full shape (FS) galaxy survey data in combination with CMB to constrain the parameters.
More specifically, we employ  the multipoles of galaxy power spectra data from Baryonic Oscillation Spectroscopic Survey (BOSS) Data Release 12 (DR12),  which has been combined into a full shape (FS) likelihood in \cite{Philcox:2020vbm,Philcox:2020vvt,Philcox:2021kcw}. As demonstrated in the present article, these early universe neutrino interactions impact the galaxy power spectra multipoles differently depending on the interaction redshift. Our investigations suggest improved bounds on interaction redshift ($z_{\rm int}$)  over those obtained from (Planck+ BAO) only analysis \cite{Kreisch:2019yzn,Kreisch:2022zxp,Taule:2022jrz} for different values of  $n_{\rm int}$. The $n_{\rm int}=5$ model  admits a concrete particle physics model \cite{Kreisch:2019yzn,Kreisch:2022zxp,RoyChoudhury:2019hls,RoyChoudhury:2020dmd,Das:2020xke}, thereby allowing us to place better constraints on the effective coupling $G_{\rm eff}$. We also examine forecast results for future LSS observation, like Euclid mission in a joint analysis with CMB-S4 and Planck  baseline, providing further improvement in relevant bounds. More details on the improved bounds are presented in Sec.~\ref{sec:results_analysis} and \ref{sec:forecast} and also highlighted in the Conclusion section.   

The paper is organized as follows. In Sec.~\ref{sec:neutrinos_CPT}, we briefly review the basics of cosmological perturbation theory (CPT) in presence of massive neutrinos within the linear regime in presence of interactions and modeling neutrino interactions in the early universe. Following that, in Sec.~\ref{sec:galaxy_ps} we discuss how these scenarios affect the galaxy power spectra in mildly non-linear regime. The data and methodology used in this paper are presented in Sec.~\ref{sec: data&methodology}, while our results and discussions are shown in Sec.~\ref{sec:results_analysis}. Further in Sec.~\ref{sec:forecast}, we  present the forecast for future missions and finally we summarize in Sec.~\ref{sec:summary}. The detailed (6+2) parameter posterior distributions for all the cases are presented in Appendix~\ref{appendix:a} and \ref{appendix:b}.

\section{Neutrinos in cosmological perturbations}
\label{sec:neutrinos_CPT}
\subsection{Neutrino perturbation equations}
\label{subsec:pert_equn}

In the primordial universe, neutrinos, as relativistic entities, generate anisotropic stress within the perturbed Einstein metric. This anisotropy arises from the velocity perturbations in the fluid equations of free-streaming neutrinos and plays a pivotal role in shaping the CMB spectrum. Such anisotropic stress drives the metric perturbation that suppresses the CMB angular power spectra. Additionally, the rapid free-streaming of neutrinos at nearly the speed of light during this epoch introduces a phase shift in BAO. On the other hand, presence of interactions among neutrinos can delay the onset of free-streaming by dampening the anisotropic stress, thereby altering the dynamics. The Boltzmann hierarchy equations, accounting for neutrino interactions, can be expressed as follows \cite{Ma:1995ey, Kreisch:2019yzn},

\begin{subequations}
\label{Boltzmannhierarchy}
\begin{align}
     \frac{d\Psi_0}{d\tau} &= -{qk\over \epsilon}\Psi_1+{1\over 6}\dot{h} {d\ln f_0\over d\ln q}\,, \label{eq:psi0} \\
     \frac{d\Psi_1}{d\tau}  &= {qk\over 3\epsilon} \left(\Psi_0- 2 \Psi_2 \right) \,, \label{eq:psi1} \\
     \frac{d\Psi_2}{d\tau}  &= {qk\over 5\epsilon} \left(2\Psi_1 - 3\Psi_3 \right) - \left( {1\over15}\dot{h} + {2\over5} \dot{\eta} \right)
	{d\ln f_0\over d\ln q} - a \,\Gamma_{\nu} \,\Psi_2\,,   \label{eq:psi2}\\
    \frac{d\Psi_l}{d\tau} &= {qk \over (2l+1)\epsilon} \left[ l\Psi_{l-1} - (l+1)\Psi_{l+1} \right]- a \,\Gamma_{\nu}\, \Psi_l   \,, \quad l \geq 3 \,. \label{eq:psil}
\end{align}
\end{subequations}

Here, $f_0$ represents the background Fermi-Dirac distribution function, while $\Psi_l(k,q,\tau)$ denotes the $l^{\rm th}$  order perturbation to the distribution function, corresponding to the $l^{\rm th}$ order Legendre polynomial in the Fourier space. The comoving energy density of relativistic neutrinos is given by $\epsilon$ $(\epsilon = \sqrt{q^2+a^2 m_\nu^2})$, where $q$ is the corresponding amplitude of the comoving momentum.  Here Boltzmann hierarchy equations are expressed in synchronous gauge, where $h$ and $\eta$ represent the standard metric perturbations in this gauge. 

Additionally, $\Gamma_{\nu}$ signifies the neutrino interaction rate in the early universe, with the parameterization of this interaction rate detailed in the subsequent section. Due to the conservation of mass and momentum, the evolution equations for $\Psi_0$ and $\Psi_1$ are unaffected by the interaction terms while neutrino interactions begin to influence the Boltzmann hierarchy starting from $l = 2$.

We have implemented Eqs.~(\ref{eq:psi0}-\ref{eq:psil}) in the cosmological code \texttt{CLASS-PT} \cite {Chudaykin:2020aoj} which is an extension to the Boltzmann solver code \texttt{CLASS} \cite{lesgourgues2011cosmic,blas2011cosmic}.
The code outputs are based on the EFT of LSS \cite{Baumann:2010tm,Carrasco:2012cv,Carrasco:2013mua,Senatore:2014via,Vlah:2015zda,Senatore:2017pbn,Ivanov:2018gjr,Senatore:2014eva,Senatore:2014vja,Cabass:2022avo} as detailed in Sec.~\ref{sec:galaxy_ps}. Within \texttt{CLASS-PT}, the Boltzmann hierarchy equations have been computed in synchronous gauge following \cite{Taule:2022jrz} as usually done in standard \texttt{CLASS} code \cite{lesgourgues2011cosmic,blas2011cosmic}. For implementing the interaction scenario, we use the standard relaxation time approximation. We assume that the interaction rate $\Gamma_\nu$ only depends on the neutrino temperature and independent of the internal momenta and cosmological scales. Although the effects of including the momentum dependency of the interaction rate has been studied before in \cite{Oldengott:2014qra,Chen:2022idm}, as long as neutrino mass is negligible and we consider $\Gamma_\nu$ to be the average rate at which the neutrino free-streaming is damped, this is a good approximation to proceed. The approximations used for the neutrino Boltzmann hierarchy is mentioned in the footnote \footnote{We have used the default fluid approximation for non-cold relics \textit{i.e. CLASS-FA} following \cite{blas2011cosmic,lesgourgues2011cosmic}. Full Boltzmann hierarchy is employed until default value of $k\tau$. Also truncation of the full Boltzmann hierarchy has been done at $l_{\rm max}=17$. Additionally we consider three degenerate neutrinos and solve the hierarchy for one neutrino species. }.

\subsection{Modeling early universe neutrino interactions}
\label{subsec:early_universe_int}

Neutrino interactions across the evolutionary timeline of our universe can be depicted through a generic, model-independent framework. The interaction rates within this paradigm are articulated in terms of the Hubble parameter, with a dependency on both the interaction redshift and the neutrino temperature index. 

Neutrino interactions can be broadly categorized into those occurring in the early universe and those in the late universe. In the early universe, neutrinos are coupled through four-Fermi weak interactions up to a certain redshift, beyond which they decouple from the primordial plasma. Although even after weak interaction domination phase, they can remain coupled through self-interactions motivated from BSM physics. Conversely, in the late universe, interactions may arise from various mechanisms, including neutrino decay, self-interactions mediated by light particles, and neutrinophilic interactions.

In this model-independent framework, these interactions can be parameterized  as \cite{Taule:2022jrz}:

\begin{align}\label{eq:Gamma_powerlaw}
\Gamma_{\nu}(z,z_{\rm int}) = H(z_{\rm int}) \left(\frac{1+z}{1+z_{\rm int}} \right)^{n_{\rm int}},
\end{align}

Here, $z_{\rm int}$ represents the redshift at which the Hubble parameter equals the interaction rate, \textit{i.e.}, $\Gamma_{\nu}(z_{\rm int})= H(z_{\rm int})$. The interaction types are characterized by the power-law index $n_{\rm int}$, with $n_{\rm int} = [3,\,4,\,5]$ in the early universe and $n_{\rm int} = [-5,\,-3,\,-1,\,1]$ in the late universe. Apart from that, there are interactions which are transient in nature characterized by a free-streaming window $2000<z_{\rm int}<10^5$, see \cite{Taule:2022jrz}. A majority of such interactions are phenomenological in nature. For example, $n_{\rm int}=-5$ corresponds to the neutrino decay scenario \cite{Barenboim:2020vrr,Chen:2022idm} and $n_{\rm int}=1 $ corresponds to the case where neutrinos and anti-neutrinos annihilate to massless bosons \cite{Oldengott:2014qra, Forastieri:2019cuf}.

\begin{figure}[ht!]
    \centering
    \subfloat{\includegraphics[width=0.50\textwidth]{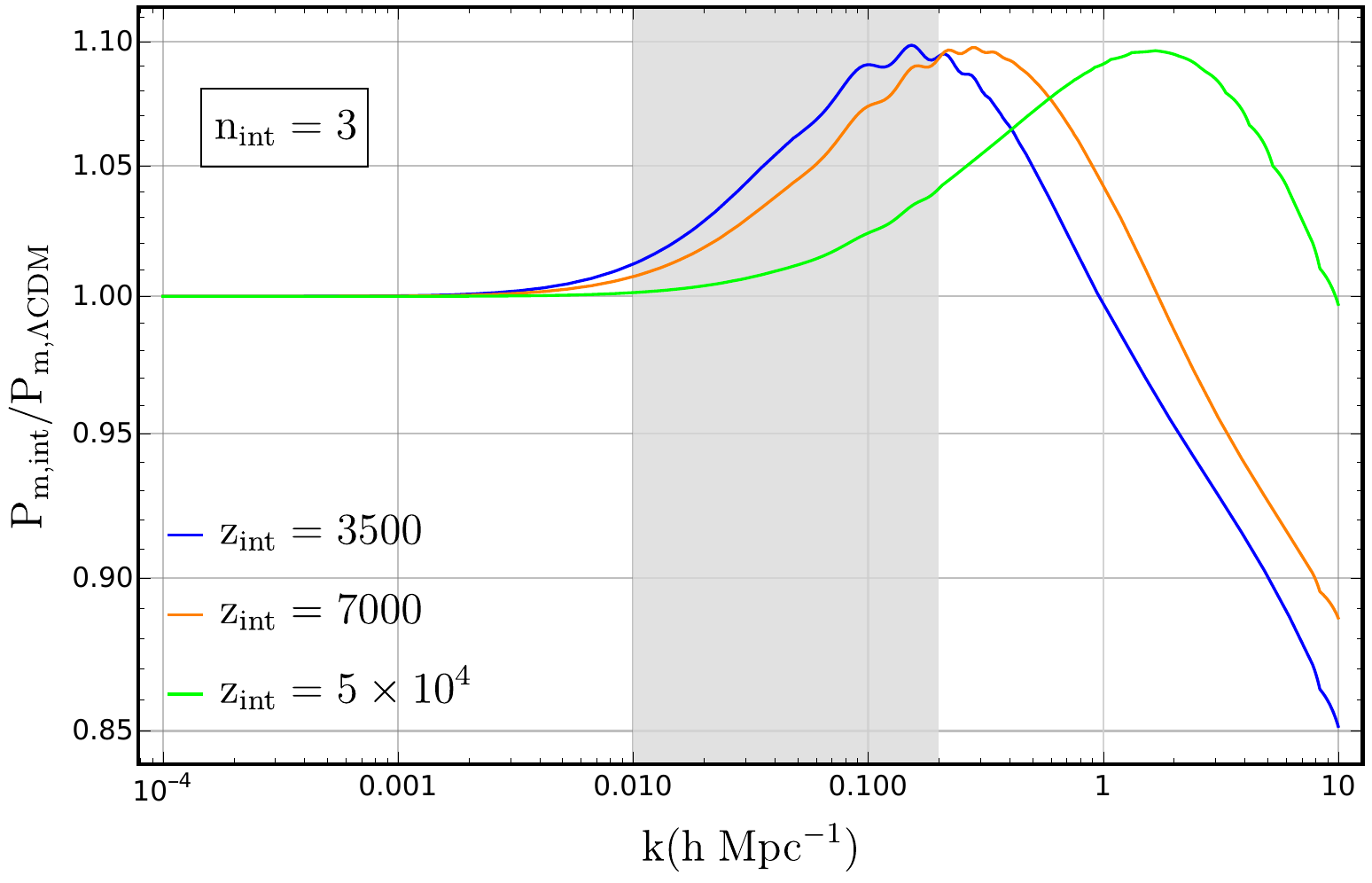} }
    \subfloat{\includegraphics[width=0.50\textwidth]{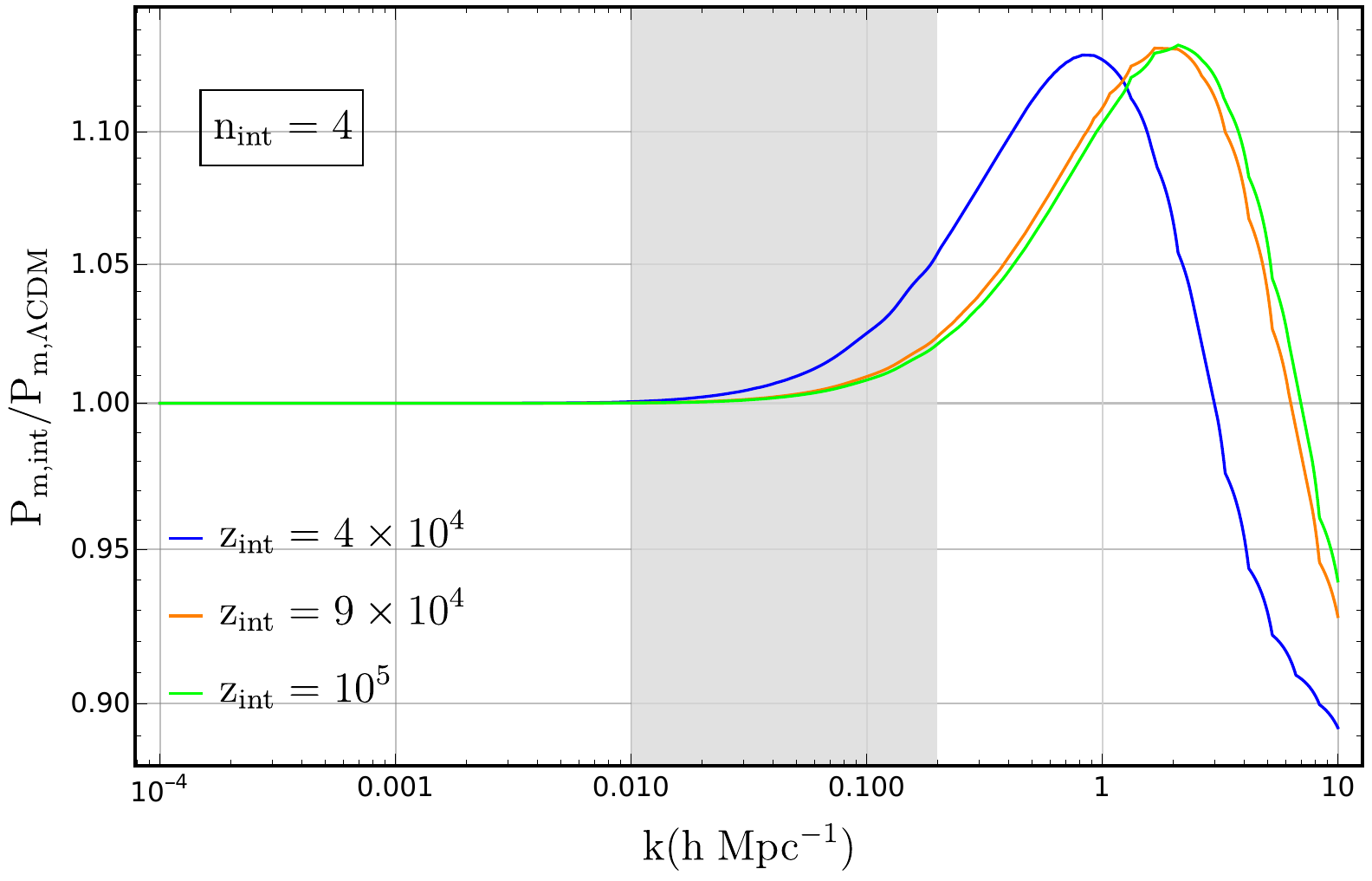} }
    \qquad
   \subfloat{\includegraphics[width= 0.50\textwidth]{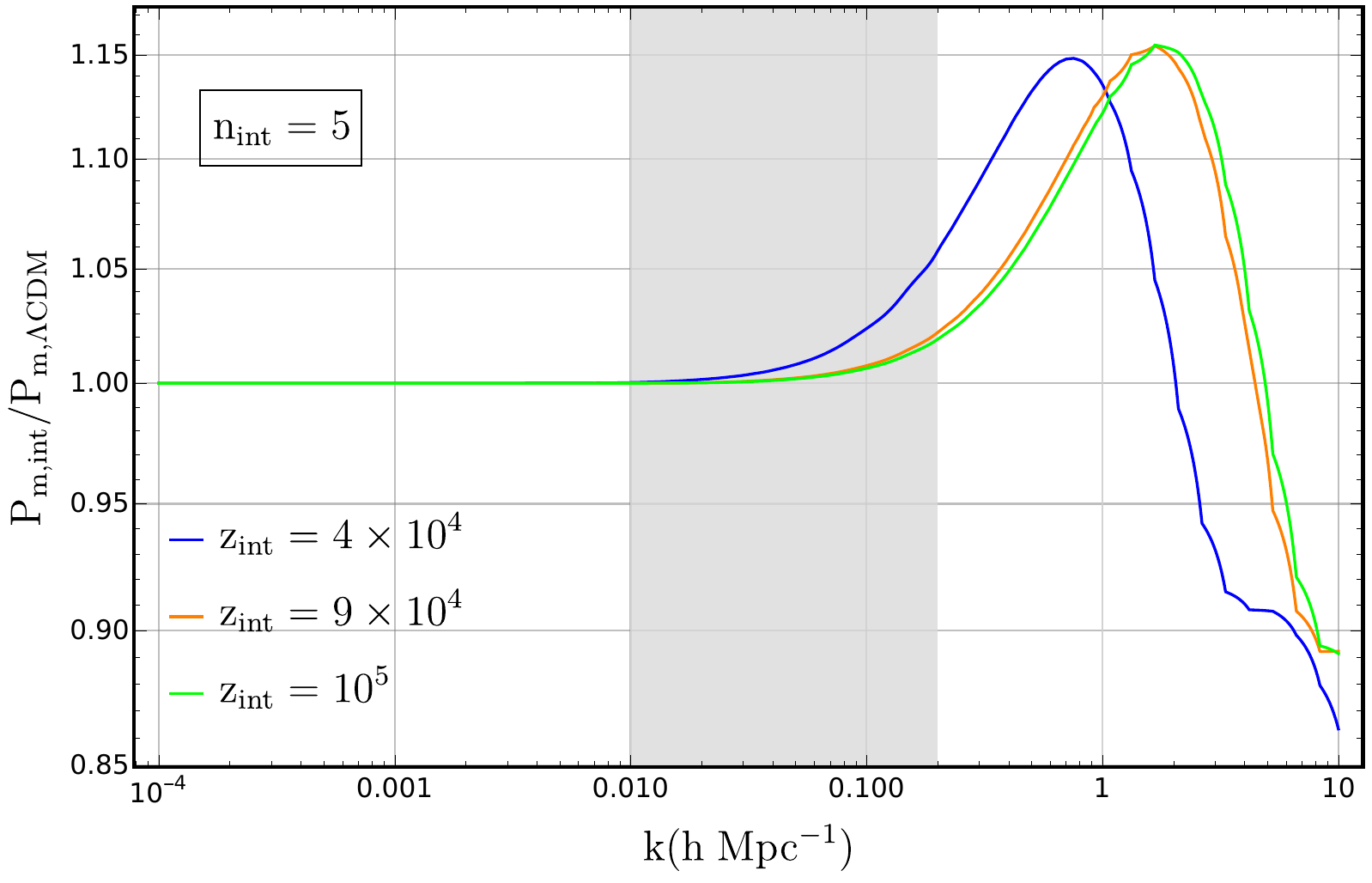} }
    \caption{\it The ratio of linear matter spectra in presence of neutrino interactions with $\Lambda$CDM + $\sum m_\nu\,$ model. The interacting models shown here correspond to $n_{\rm int}=3,4$ and 5 respectively with different interaction redshifts.}
    \label{fig:suppresion}
\end{figure}

On the other hand, $n_{\rm int}=5$ case represents a well-known particle physics scenario that can be easily mapped to neutrino self-interactions mediated by a heavy scalar mediator. There are a plethora of studies in the literature on neutrino self-interactions, both in the context of CMB \cite{Kreisch:2019yzn, Park:2019ibn, Das:2020xke, Brinckmann:2020bcn} and LSS \cite{ Camarena:2023cku, Camarena:2024zck, He:2023oke}. In particular, in our parameterization, $n_{\rm int}=5$ corresponds to the moderately interacting (MI) mode in the self-interaction models \cite{Kreisch:2019yzn, Park:2019ibn, Das:2020xke, Brinckmann:2020bcn,Camarena:2023cku, Camarena:2024zck, He:2023oke}. It can be identified whether the interactions are active at early epoch or late universe through the Hubble parameter. In the radiation domination epoch $H(z) \propto T^2$ and in the matter domination epoch $H(z) \propto T^{3/2}$, where $T$ is the background temperature. This suggests that interactions with $n_{\rm int} =4$ and 5 are dominant in the early universe. On the other hand, for $n_{\rm int}=3$, the term $\Gamma_\nu/H(z)$ is almost constant throughout the evolution history. As of now, the literature is insufficient to demonstrate if there is any obvious mapping to any specific particle physics model for $n_{\rm int} = 3$ and $4$ cases. However, they are interesting cases to explore in a model-independent framework, given their prospects in the early universe scenario.

In this article, we primarily focus on the early universe neutrino interactions, i.e. $n_{\rm int} \in [3, \, 4, \, 5] $. As pointed out in \cite{Taule:2022jrz}, interactions at low temperature as well as those that are transient in nature do not affect the matter power spectra in mildly non-linear regime. As mentioned earlier, even in absence of particle physics mapping of $n_{\rm int} =3 , 4$, these kinds of interactions do have significant effects on matter power spectra in mildly non-linear regime and hence in galaxy power spectra. It is evident from Fig.~\ref{fig:suppresion}, that neutrino interaction with $\Gamma \propto T_{\nu}^3$, enhances the matter power spectra $\sim 10$\% at scale $k \approx 0.2 \, h/{\rm Mpc}$ depending on the interaction redshifts. Also interactions with $ n_{\mathrm{int}} = 4 $ and 5 modifies the matter power spectra up to $14-15 \%$ at scales $k \approx 1\, h/{\rm Mpc}$. Since interacting neutrinos in early universe significantly affect the matter power spectra in mildly non-linear regime, this in turn have imprints on the multipoles of galaxy power spectra, discussed in detail in Sec.~\ref{sec:galaxy_ps}. 

\begin{figure}[ht!]
    \centering
    \subfloat[]{\includegraphics[width=0.50\textwidth]{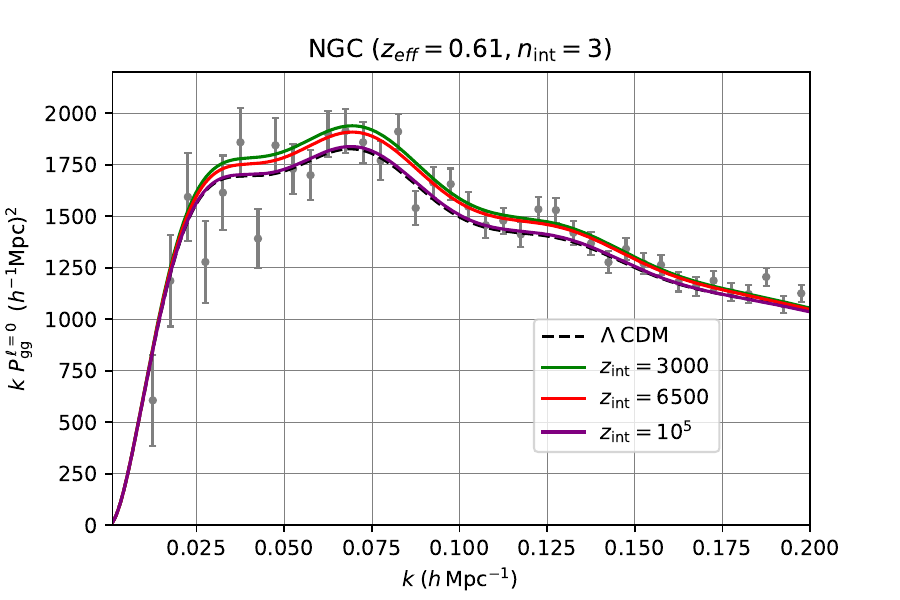} }
    \subfloat[]{\includegraphics[width=0.50\textwidth]{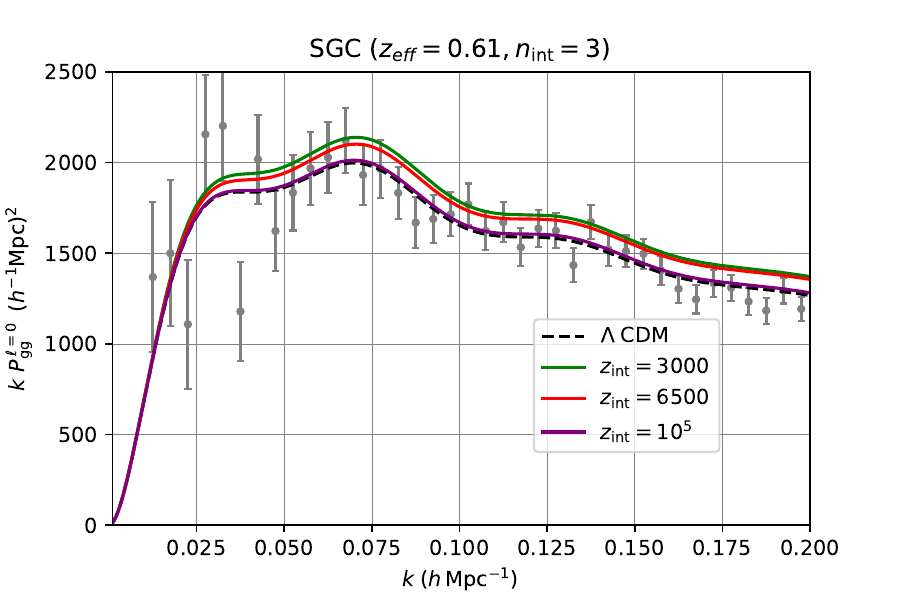} }
    \qquad
    \subfloat[]{\includegraphics[width= 0.50\textwidth]{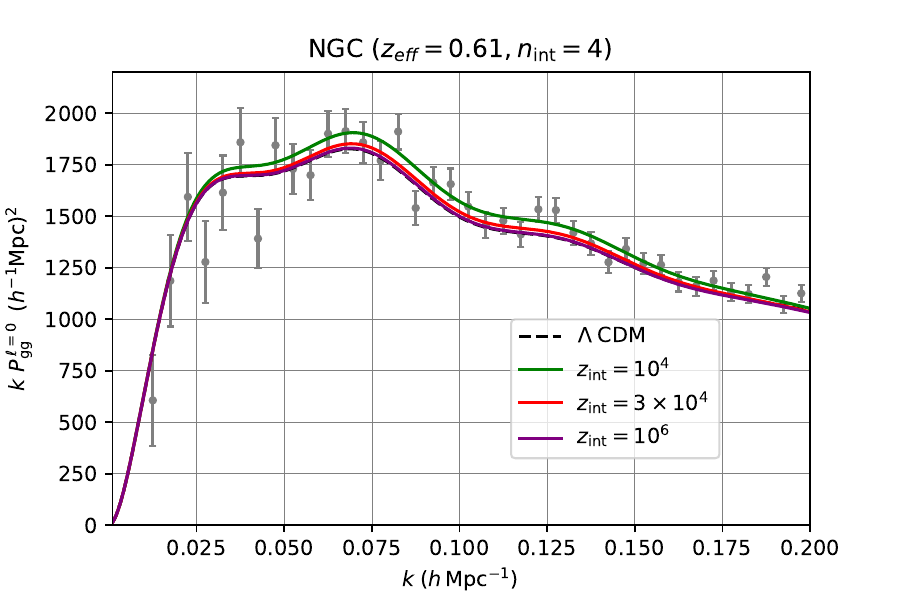} }
    \subfloat[]{\includegraphics[width=0.50\textwidth]{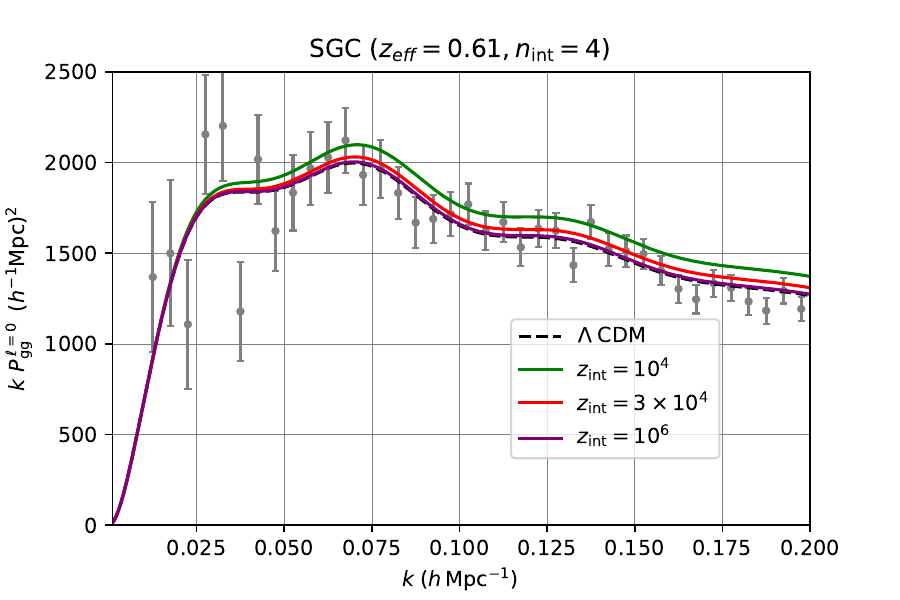} }
    \qquad
    \subfloat[]{\includegraphics[width= 0.50\textwidth]{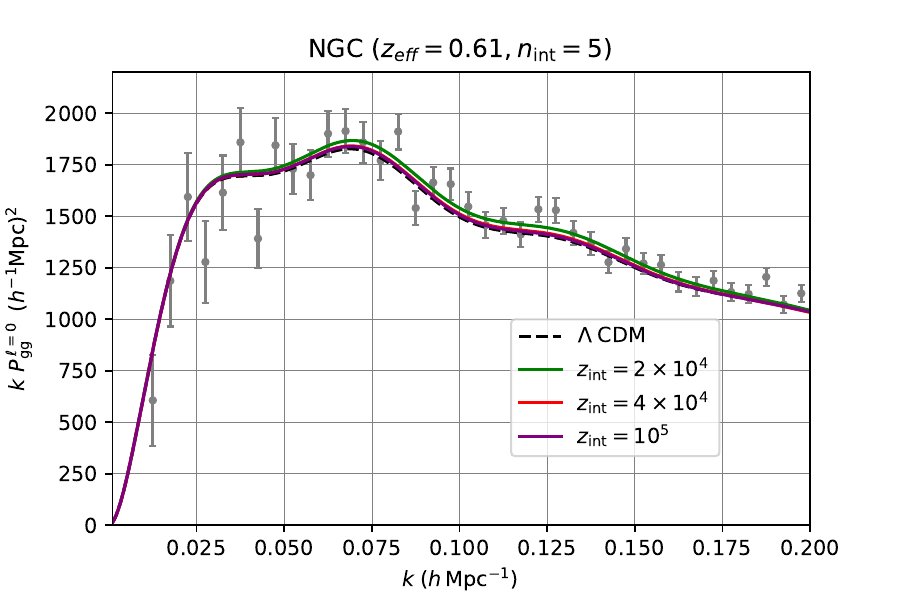} }
    \subfloat[]{\includegraphics[width=0.50\textwidth]{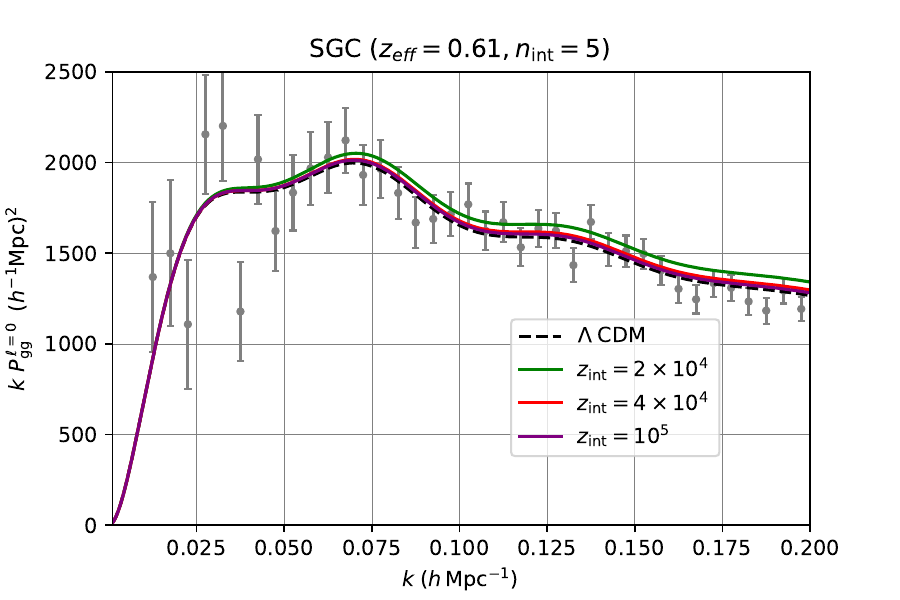} }    
    \caption{\it The plots depict the galaxy monopole spectra for $\Lambda$CDM and interacting neutrino models. Here dotted black curve represents $\Lambda$CDM best fit model and the others correspond to the effects of different interaction redshift. The \textbf{left panel} shows the effect of varying $z_{\rm int}$ on NGC CMASS galaxy samples with effective redshift $z_{\rm eff}=0.61$ for different $n_{\rm int}$ cases. In the \textbf{right panel} the same effects are shown for SGC CMASS galaxy samples for varying $n_{\rm int}$. The error bar shown here are generated with publicly available BOSS DR12 full shape spectra likelihood following \cite{Philcox:2021kcw}.}
    \label{fig:galaxy_multipoles}
\end{figure}

\section{Effect of neutrino interactions on full shape galaxy power spectra}
\label{sec:galaxy_ps}
 
The effects of neutrinos are most effectively analyzed through the study of the evolution of  gravitational potential. Within the standard framework, neutrino anisotropic stress is the primary factor contributing to the difference in the gravitational potentials, as described by the perturbed Einstein equation,
\begin{equation}\label{eq:anisotropic}
     k^2 (\phi - \psi) = 16 \pi G a^2 \rho_{\rm tot}R_\nu \sigma_\nu
\end{equation}
where $\rho_{\rm tot}$ represents the total radiation energy density, and $R_\nu$ is the fractional energy density of free-streaming neutrinos. In the standard $\Lambda$CDM model, free-streaming neutrinos make up approximately $41 \%$ of the total radiation energy density \cite{Lesgourgues:2013sjj,Bashinsky:2003tk,Dolgov:2002wy}. The difference in the evolution of the potentials in Eq.~(\ref{eq:anisotropic}) affects the growth of matter fluctuations in subsequent evolution history of our universe.

Neutrino interactions modify the difference in the gravitational potentials, by suppressing the anisotropic stress term $\sigma_\nu$, leading to  $\phi-\psi \approx 0$, thereby affecting the growth of dark matter fluctuations in different scales. Large scale modes enter the horizon well after neutrino free-streaming and hence remain unaffected by these interactions, evolving as in standard $\Lambda$CDM cosmology.

On the contrary, modes with $k \sim 10 \, h/{\rm Mpc}$, that enter the horizon while neutrinos are still tightly coupled to the primordial plasma through the corresponding interactions, experience an initial amplitude enhancement due to the amplification of the gravitational potential. However, the absence of anisotropic stress also amplifies the magnitude of oscillations in the gravitational potential for these small scales in comparison to $\Lambda$CDM scenario. As a result, $\psi$ decays slowly for these modes than $\Lambda$CDM paradigm. This results in a damping of dark matter fluctuations and a suppression of the matter power spectrum at these scales. 

Modes of particular interest are those with $k\sim 0.1 \, h/{\rm Mpc}$, which enter the horizon as neutrinos begin to free-stream. While these modes also experience an initial enhancement, the gravitational potential rapidly decays to the $\Lambda$CDM  baseline, leading to a subsequent enhancement in the matter power spectrum. The interplay of these effects produces a bump-like feature in the matter power spectrum in the mildly non-linear regime.

As illustrated in Fig.~\ref{fig:suppresion}, the onset of free-streaming determined by the redshift of decoupling, significantly influences the matter power spectrum by introducing distinctive features that depend on the nature of the interactions involved. In all the plots in Fig.~\ref{fig:suppresion}, standard cosmological parameters are fixed to the best-fit $\Lambda$CDM values and $\sum m_\nu$ is fixed to $0.12 \, {\rm eV}$.
Also all the power spectra are plotted at redshift 0.61, as probed by BOSS DR12. Specifically, for interactions that scale as $T_\nu^3$, the matter power spectrum exhibits an enhancement of roughly 10$\%$ over the wavenumber range $k \sim [0.1, 10]\, h/{\rm Mpc}$ for interaction redshifts $z_{\rm}= 3500, \, 7000$ and $5 \times 10^4$.
This bump-like feature  shifts toward smaller scales as the redshift of decoupling increases, reflecting the earlier transition to free-streaming. This shift is similarly observed for interactions characterized by $n_{\rm int} = 4$ and $n_{\rm int} = 5$, where the enhancement in the matter power spectrum becomes even more pronounced. In particular, for $n_{\rm int} = 4$, the enhancement reaches nearly 14$\%$ for interaction redshifts within $10^4$-$10^5$ ranges, and the associated feature is also displaced towards smaller scales as shown in Fig.~\ref{fig:suppresion}  demonstrating a clear correlation between the interaction strength and the scale of the enhancement. The interaction most relevant from the perspective of particle physics, where the interaction rate scales as $\Gamma_\nu \propto T_\nu^5$, leads to a substantial enhancement in the matter power spectrum, approaching 15$\%$ for interaction redshifts within $10^4$-$10^5$ ranges in the scales that are critical for galaxy surveys as shown in Fig.~\ref{fig:suppresion}. This demonstrates that the redshift at which neutrinos begin to free-stream has non-trivial effects on the matter power spectrum in both linear and mildly non-linear regimes.

In the standard $\Lambda$CDM framework, massive neutrinos begin to free-stream  after they become non-relativistic, deep within the matter-dominated epoch. The free-streaming scale in $\Lambda$CDM cosmology is given by $k_{\rm FS}(z_{\rm NR}) \approx 0.018 \, \sqrt{\Omega_{\rm m} ( m_\nu/1 \, \text{eV})} \, h/{\rm Mpc}$ \cite{Lesgourgues:2013sjj, Levi:2016tlf,Pal:2023dcs}, where $z_{\rm NR}$ is the non-relativistic transition redshift. The suppression of the matter power spectrum by massive neutrinos at scales much greater than the free-streaming scale ($k \gg k_{\rm FS}$) and their CDM-like behavior on larger scales ($k \ll k_{\rm FS}$) remain unaffected by early universe interactions, which persist until matter-radiation equality without altering the matter-dominated epoch.

As already mentioned, we have incorporated neutrino interactions into the \texttt{CLASS-PT} \cite{Chudaykin:2020aoj} code. The linear perturbation theory remains valid up to modes with $k \lesssim 0.1\, h/{\rm Mpc}$, beyond which \texttt{CLASS-PT} code applies one-loop corrections to the matter power spectra up to mildly non-linear scales. These one-loop corrections are based on the EFT of LSS in Eulerian space, using  EDS (Einstein De-Sitter) convolution kernels approximation \cite{Bernardeau:2001qr}. Additionally, the galaxy power spectra multipoles are computed with redshift space distortion (RSD) corrections. We can use the EDS approximation for the dark matter sector even in the presence of neutrino interactions, since massive neutrinos are known to free-stream in the matter dominated epoch \cite{Camarena:2023cku,He:2023oke}. The one-loop corrected redshift space galaxy power spectra using EFT of LSS become unreliable for modes  $k \gtrsim 0.25\, h/{\rm Mpc}$ \cite{Senatore:2014vja,Ivanov:2019pdj,DAmico:2019fhj,Chudaykin:2020aoj}, thus our analysis is limited to $k_{\rm max} \approx 0.2\, h/{\rm Mpc}$. In contrast, the real space power spectrum is reliable up to scales $k_{\rm max} \approx 0.4\, h/{\rm Mpc}$ \cite{Ivanov:2021fbu}. 

In Fig.~\ref{fig:galaxy_multipoles} we have shown the effects of free-streaming redshift ($z_{\rm int}$) on galaxy monopoles only (since the effects on galaxy quadrupole moments are not so prominent). The data points and error bars in all the plots are derived from BOSS DR12 galaxy full shape power spectra likelihood \footnote{\href{https://github.com/oliverphilcox/full_shape_likelihoods}{https://github.com/oliverphilcox/full_shape_likelihoods}}, as detailed in \cite{Philcox:2021kcw}. In the left panels of Fig.~\ref{fig:galaxy_multipoles}, the data points and error bars of monopoles are extracted from the North Galactic Cap (NGC) data chunk over the redshift bin $0.5<z<0.75$ with effective redshift $z_{\rm eff}=0.61$, while the right panel shows similar data for the South Galactic Cap (SGC).
In all the plots of Fig.~\ref{fig:galaxy_multipoles}, the black dotted line represents $\Lambda$CDM case. We observe that the interaction parameters $z_{\rm int}$ and $n_{\rm int}$ significantly influence the galaxy monopoles compared to the $\Lambda$CDM case. For all the figures, the best-fit values considered for the EFT nuisance parameters are taken assuming background $\Lambda$CDM cosmology, just to demonstrate the effect of $z_{\rm int}$ on the galaxy power spectra.

\section{Current data and methodology}
\label{sec: data&methodology}

As previously mentioned, we utilize a modified version of \texttt{CLASS-PT}\footnote{\href{https://github.com/Michalychforever/CLASS-PT}{https://github.com/Michalychforever/CLASS-PT}} \cite{Chudaykin:2020aoj} to perform the Bayesian analysis of the model in constraining the model parameters $z_{\rm int}$ and $\sum m_\nu$, as well as the standard cosmological parameters, using the latest version of the Markov Chain Monte Carlo (MCMC) sampler, \texttt{MontePython}\footnote{\href{https://github.com/brinckmann/montepython_public}{https://github.com/brinckmann/montepython_public}} 
\cite{audren2013conservative,Brinckmann:2018cvx}. For our analysis, we consider the combinations of the following currently available datasets:

\begin{enumerate}[*]
    \item \textit{CMB:} low-$\ell$ and high-$\ell$ CMB temperature power spectrum and low-$\ell$ and high-$\ell$ CMB E mode polarization and their temperature cross correlation from Planck \cite{Planck:2018vyg}.
    \item \textit{BAO:} On top of Baryonic Oscillation Spectroscopic Survey (BOSS) DR12 BAO, we used BAO data from Lyman-$\alpha$ (Ly$\alpha$) absorption and quasars at an effective redshift, $z_{\rm eff} = 2.33 $ from DR16 extended BOSS (eBOSS) survey \cite{eBOSS:2019dcv,eBOSS:2020tmo}, which we denote as \underline{BAO} throughout our analysis.
    
    \item \textit{Galaxy Full Shape Spectra:} We utilize the dataset from the twelfth data release of BOSS DR12 \cite{eisenstein2011sdss, dawson2012baryon, BOSS:2016wmc} and its corresponding window-free galaxy power spectrum \cite{Philcox:2020vbm, Philcox:2021kcw} to investigate potential new interactions in the neutrino sector. Since eBOSS DR16 datasets have yet not been combined into a full shape likelihood, we have considered the latest full shape power spectra, {\it i.e}., DR12 for our analysis.  The BOSS DR12 galaxies are divided into four subsets, corresponding to two redshift slices $0.2 < z < 0.5$ from the LOWZ sample (effective redshift, $z_{\rm eff} = 0.38$) and $0.5 < z < 0.75$ from the CMASS sample ($z_{\rm eff} = 0.61$) and two sky cuts in the North and South Galactic Caps (NGC and SGC). The galaxy power spectrum data is provided for each subset. We denote the combined dataset as \underline{Full Shape (FS)} throughout the analysis.
    
    To explore a possible delay in the onset of neutrino free-streaming, we analyze the multipoles of the galaxy power spectrum $P_\ell(k, z)$ ($\ell = 0, 2, 4$) \cite{Philcox:2021kcw} along with the $Q_0(k, z)$ estimator \cite{Ivanov:2021fbu}, which is closely related to the real-space power spectrum and derived using a linear combination of the first few power spectrum multipoles. For the reason mentioned earlier, our primary analysis conservatively uses the multipoles within the wavenumber range $k_{\rm min} = 0.01$ $h/{\rm Mpc}$ to $k_{\rm max} = 0.2$ $h/{\rm Mpc}$ for redshift-space. Since real-space perturbation theory is applicable to smaller scales, we also consider measurements of the $Q_0$ estimator in the range $k_{\rm min} = 0.2$ $h/{\rm Mpc}$ to $k_{\rm max} = 0.4$ $h/{\rm Mpc}$. In both cases, we use a bin width of $\Delta k = 0.005$ $h/{\rm Mpc}$. Additionally, we utilize the reconstructed power spectrum to provide constraints on the Alcock-Paczynski (AP) parameters \cite{Philcox:2020vvt}.

    Our analysis employs the BOSS likelihood \cite{Philcox:2021kcw}, which analytically marginalizes over the nuisance parameters that enter linearly into the power spectrum, such as the counterterms (monopole $c_0$, quadrupole $c_2$, hexadecapole $c_4$, and fingers-of-God $\tilde{c}$), the third order galaxy bias $b_{\Gamma_3}$, and the stochastic contributions ($P_{\rm shot}$, $a_0$, and $a_1$). The covariance matrix used for this likelihood is computed using MultiDark-Patchy 2048 simulations \cite{Kitaura:2015uqa, Rodriguez-Torres:2015vqa}.
     \end{enumerate}

With these datasets, we run MCMC code MontePython with the following free parameters: the 6 standard cosmological parameters for $\Lambda$CDM, namely: CDM density $\omega_{\rm cdm}$, baryon density $\omega_{\rm b}$, angular scale of the sound horizon at recombination $\theta_{s}$, the amplitude $A_s$,  spectral index 
of the primordial spectra  $n_{s}$ and finally, the optical depth to reionization $\tau_{ reio}$. Additionally, our model parameters include the interaction redshift $z_{\rm int}$ and the sum of neutrino mass $\sum m_\nu$\footnote{We have fixed the value of $N_{\rm eff}$ to 3.044 \cite{Bennett:2020zkv,Drewes:2024wbw}} in our analysis..
In Table~\ref{tab:prior}, prior ranges of the cosmological and model parameters, used for the present analysis are listed.  

\begin{table}[!ht]
    \centering
    \begin{tabular}{|c|c|}
    \hline 
    \hline
        Parameter & Prior \\
        \hline
        $100~\omega{}_{\rm b }$ & Flat, unbounded\\ 
        $\omega{}_{\rm cdm }$ & Flat, unbounded\\ 
        $100~\theta{}_{\rm s }$ & Flat, unbounded\\ 
        ${\rm ln}(10^{10}A_{\rm s })$ & Flat, unbounded\\ 
        $n_{s }$ & Flat, unbounded\\ 
        $\tau{}_{\rm reio }$ & Flat, $\tau{}_{\rm reio } \geq 0.004$ \\ 
        $\log_{10}z_{\rm int}$ & \,Flat,\,\,$[3,6]$\\ 
        $\sum m_\nu[{\rm eV}]$ &\,Flat,\,\,$[0.001,1.0]$\\ 
        \hline
        \hline
    \end{tabular}
    \caption{\it MCMC parameters and priors} 
    \label{tab:prior}
\end{table}

\section{Results and analysis}
\label{sec:results_analysis}

In this section, we present  the results  obtained from the above methodology using combinations of various datasets. First, let us present the constraints on the $(6+2)$ parameters using Planck TT, TE, EE + BAO datasets as well as Planck TT, TE, EE + BAO + FS datasets.  In our analysis, we focus exclusively on interactions within the neutrino sector, ensuring that the equivalence principle remains valid in the dark matter sector. The FS measurements incorporate both the non-wiggle part of the power spectra $P(k)$ and the geometrical information from the wiggle part of $P(k)$ (\textit{i.e.} BAO). Including the shape information from $P(k)$, along with the geometrical feature in the BAO data, results in tighter constraints on the cosmological as well as model parameters. 

\begin{figure}[hbt!]
    \centering
    \subfloat[]{\includegraphics[width=0.50\textwidth]{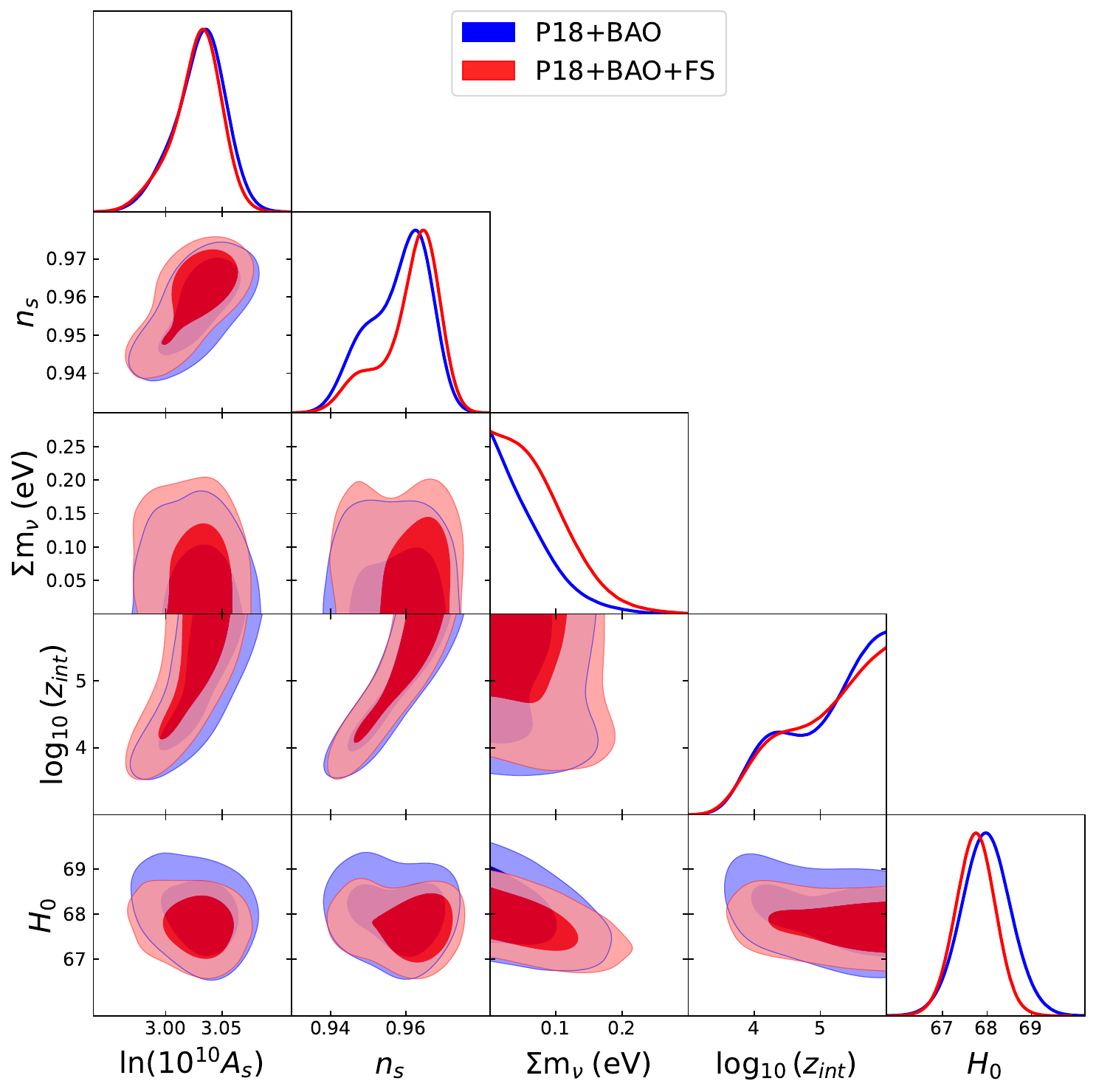} }
    \subfloat[]{\includegraphics[width=0.50\textwidth]{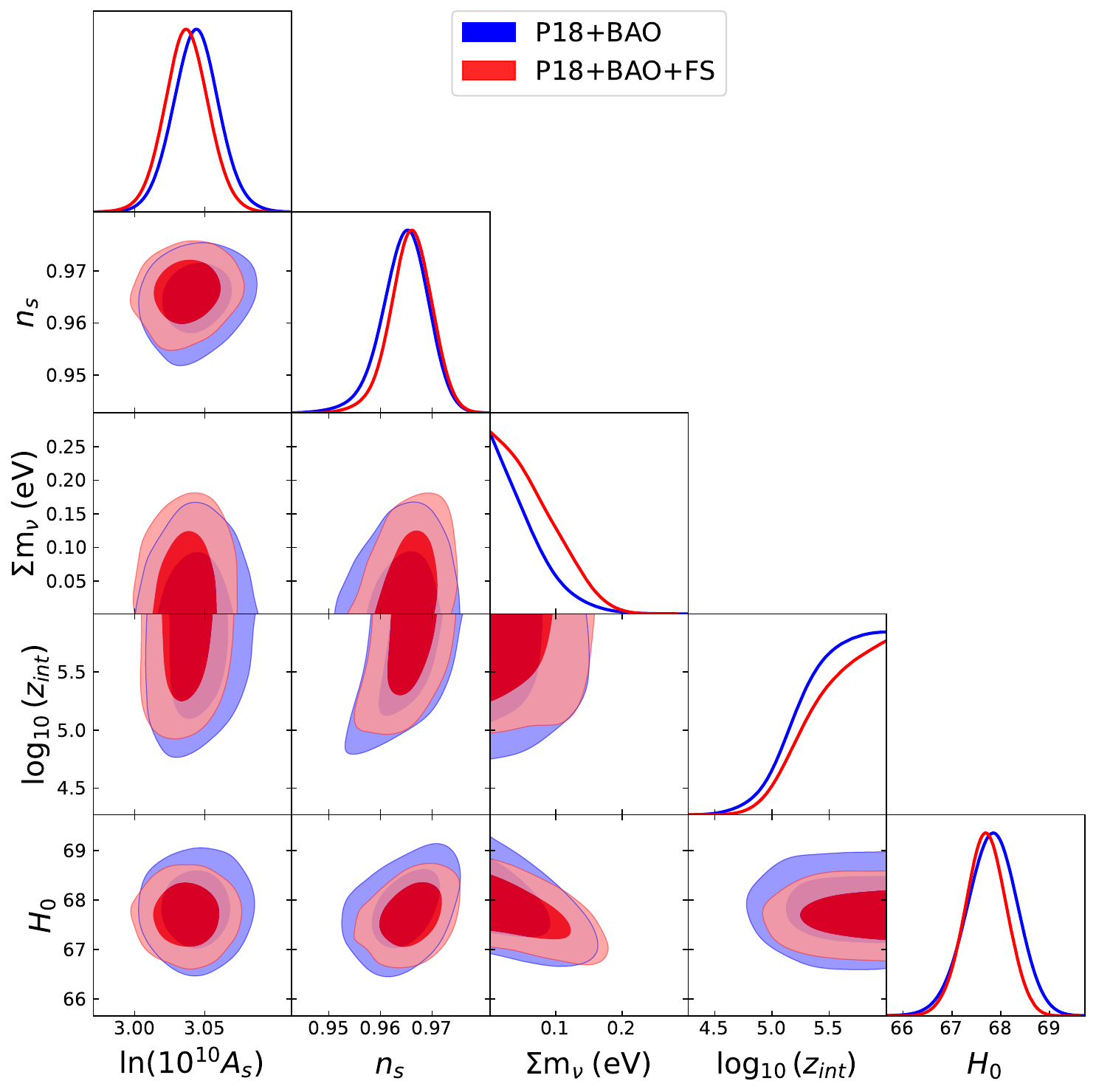} }
    \qquad
   \subfloat[]{\includegraphics[width= 0.55\textwidth]{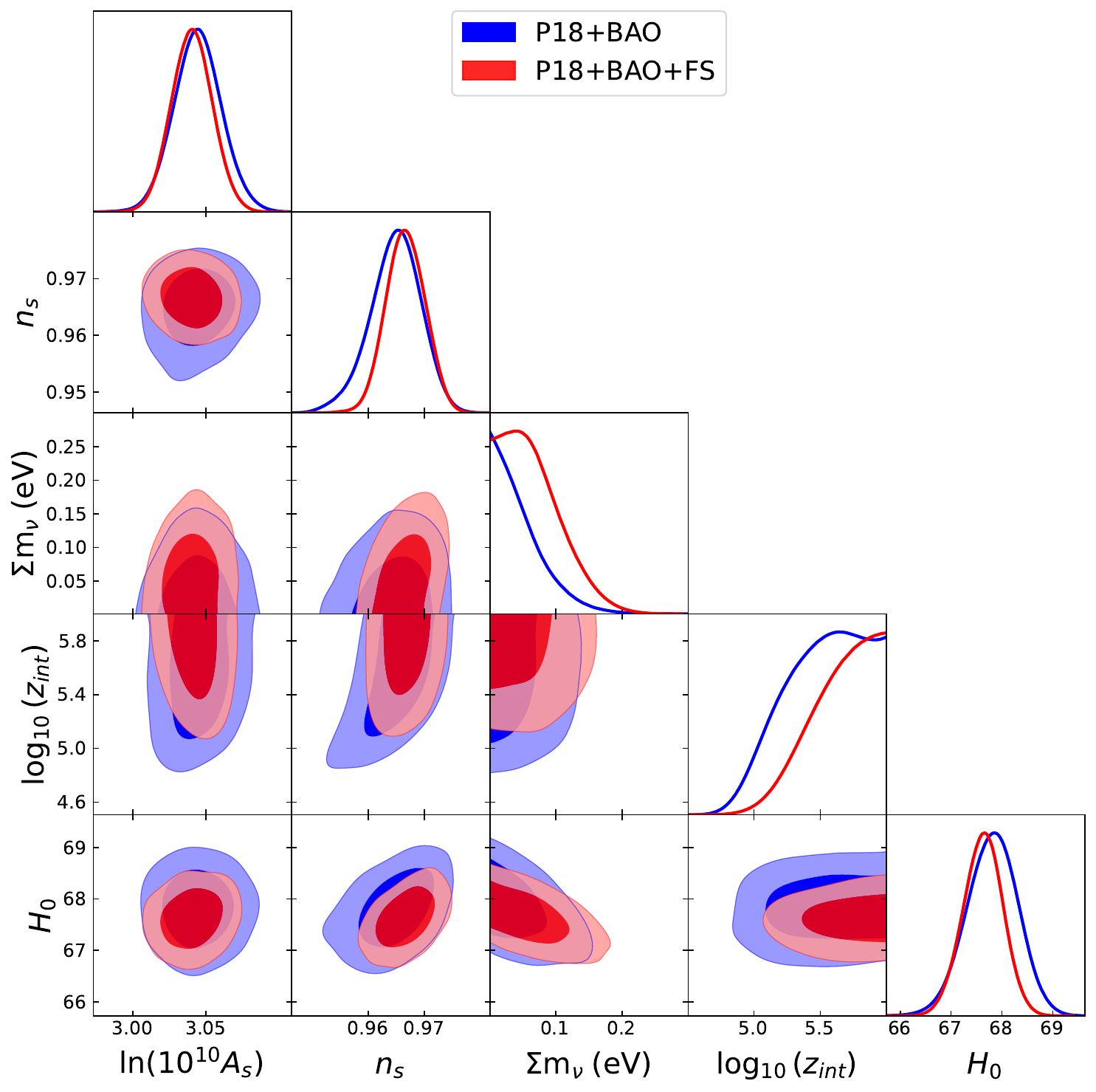} }
    \caption{\it Triangular plots representing the 2D and 1D marginalized posterior distributions for $A_s, \,n_s,\,H_0,\,\sum m_\nu$ and $z_{\rm int}$ as obtained from the \textbf{Planck+BAO} (blue) and \textbf{Planck+BAO+FS} (red) datasets. Here  the figures (a), (b) and (c) represent the $n_{\rm int}=3,\, 4$ and $5$ cases, respectively. Detailed triangular plots for each individual case have been listed in Appendix~\ref{appendix:a}.}
    \label{fig:combined_analysis}
\end{figure}

Fig.~\ref{fig:combined_analysis} illustrates the triangular plots with $1\sigma$ and $2\sigma$ confidence contours of major parameters for the three different cases ($n_{\rm int}=3$, 4 and 5) (while the posterior distributions for all the parameters for all the three cases are individually displayed in Figs.~\ref{2d_posterior_nint3}, \ref{2d_posterior_nint4} and \ref{2d_posterior_nint5} in Appendix~\ref{appendix:a}).  The corresponding parameter values are listed in Tables \ref{tab:nint3}, \ref{tab:nint4} and \ref{tab:nint5}. In Fig.~\ref{fig:combined_analysis}, the constraints on different interaction scenarios for the combined Planck TT, TE, EE + BAO dataset as well as for the Planck TT, TE, EE + BAO + FS dataset are shown in blue and red respectively. The Planck + BAO dataset (blue) yields  constraints  on all standard cosmological parameters for $n_{\rm int }= 4, 5$ cases, which fall within the $1\sigma$ bounds of vanilla $\Lambda$CDM cosmology, except for the $n_{\rm int}=3$ case, where strong degeneracies between $A_s$, $n_s$ and $z_{\rm int}$ are observed, consistent with the previous analyses \cite{Taule:2022jrz}.

\FloatBarrier

 The inclusion of the galaxy full shape (FS) dataset slightly modifies these constraints, though they remain more or less within the $1\sigma$ uncertainty of the former dataset. However, there are certain characteristic changes upon inclusion of the FS data, that need to be pointed out. 

As pointed out in previous studies \cite{Kreisch:2019yzn,Kreisch:2022zxp,Camarena:2023cku,He:2023oke}, in the strongly interacting (SI) mode of self-interacting model, the changes in the CMB anisotropy and galaxy power spectra can be absorbed into modifications of the primordial scalar power spectra, specifically a decrease in the amplitude $A_s$ and the spectral index $n_s$ \cite{Kreisch:2019yzn,Kreisch:2022zxp,Camarena:2023cku}. These strong degeneracies are absent in our analysis for $n_{\rm int}=4$ and $5$ respectively due to the modeling of the interactions, as we did not consider the explicit momentum dependencies of the interaction cross section in the Boltzmann hierarchy equations. Essentially our model corresponds to the moderately interacting (MI) mode in the self-interacting neutrino models. The error from this assumption is negligible for the purposes of our analysis, as noted in \cite{Oldengott:2017fhy}.  Although for $n_{\rm int}=3$ model, $\Gamma_\nu/H(z)$ is almost constant throughout the evolution and as a result neutrinos decouple comparatively at late time modifying $A_s, \, n_s$  which implies the degeneracy as obtained in Fig~\ref{fig:combined_analysis}.

\begin{table}[h!]
\centering
\renewcommand{\arraystretch}{1.5} 
\resizebox{\linewidth}{!}{ 
\begin{tabularx}{\textwidth}{|>{\centering\arraybackslash}p{2.2cm}|X|X|X|X|}
\hline
\multirow{2}{*}{\textbf{Parameters}} & \multicolumn{2}{c|}{Planck+BAO} & \multicolumn{2}{c|}{Planck+BAO+FS} \\
\cline{2-5}
& \textbf{Mean} $\bm{\pm \; 1 \sigma} $ & \textbf{Mean} $\bm{\pm \; 2 \sigma }$ & \textbf{Mean} $\bm{\pm \; 1 \sigma} $ & \textbf{Mean} $\bm{\pm \; 2 \sigma}$ \\
\hline
\rowcolor{gray!10}
$100~\omega_{\rm b}$ & $2.242^{+0.014}_{-0.014}$ & $2.242^{+0.027}_{-0.026}$ & $2.240^{+0.013}_{-0.013}$ & $2.240^{+0.025}_{-0.026}$ \\
$\omega_{\rm cdm}$ & $0.1197^{+0.0011}_{-0.0011}$ & $0.1197^{+0.0021}_{-0.0022}$ & $0.1194^{+0.0010}_{-0.0010}$ & $0.1194^{+0.0020}_{-0.0019}$ \\
\rowcolor{gray!10}
$100~\theta_{s}$ & $1.042^{+0.001}_{-0.001}$ & $1.042^{+0.004}_{-0.001}$ & $1.042^{+0.0012}_{-0.0004}$ & $1.042^{+0.0033}_{-0.0007}$ \\
${\rm ln}(10^{10}A_{\rm s })$ & $3.032^{+0.019}_{-0.025}$ & $3.032^{+0.038}_{-0.052}$ & $3.030^{+0.018}_{-0.024}$ & $3.030^{+0.034}_{-0.053}$ \\
\rowcolor{gray!10}
$n_{s}$ & $0.9598^{+0.0062}_{-0.0111}$ & $0.9598^{+0.0107}_{-0.0183}$ & $0.9631^{+0.0049}_{-0.0109}$ & $0.9631^{+0.0090}_{-0.0204}$ \\
$\tau_{ reio}$ & $0.0547^{+0.0077}_{-0.0070}$ & $0.0547^{+0.01599}_{-0.01363}$ & $0.0515^{+0.0075}_{-0.0074}$ & $0.0515^{+0.0158}_{-0.0150}$ \\
\rowcolor{gray!10}
$\log_{10}z_{\rm int}$ & $>4.23$ & $>3.77$ & $>4.50$ & $>3.90$ \\
$\sum m_\nu[{\rm eV}]$ & $<0.10$ & $<0.17$ & $<0.12$ & $<0.19$ \\
\rowcolor{gray!10}
$H_0$ & $67.98^{+0.54}_{-0.54}$ & $67.98^{+1.11}_{-1.10}$ & $67.74^{+0.43}_{-0.45}$ & $67.74^{+0.85}_{-0.89}$ \\
\hline
\end{tabularx}
}
\caption{\it Cosmological and model parameter constraints with mean values and \textbf{68\% and  95\% C.L.} for $n_{\rm int}=3$ model. The limits are shown with \textbf{Planck+BAO} and \textbf{Planck+BAO+FS} datasets.}
\label{tab:nint3}
\end{table}

\begin{table}[h!]
\centering
\renewcommand{\arraystretch}{1.5} 
\resizebox{\linewidth}{!}{ 
\begin{tabularx}{\textwidth}{|>{\centering\arraybackslash}p{2.2cm}|X|X|X|X|}
\hline
\multirow{2}{*}{\textbf{Parameters}} & \multicolumn{2}{c|}{Planck+BAO} & \multicolumn{2}{c|}{Planck+BAO+FS} \\
\cline{2-5}
& \textbf{Mean} $\bm{\pm \; 1 \sigma} $ & \textbf{Mean} $\bm{\pm \; 2 \sigma }$ & \textbf{Mean} $\bm{\pm \; 1 \sigma} $ & \textbf{Mean} $\bm{\pm \; 2 \sigma}$ \\
\hline
\rowcolor{gray!10}
$100~\omega_{\rm b}$ & $2.241^{+0.013}_{-0.013}$ & $2.241^{+0.027}_{-0.026}$ & $2.239^{+0.013}_{-0.012}$ & $2.239^{+0.026}_{-0.025}$ \\
$\omega_{\rm cdm}$ & $0.1194^{+0.0010}_{-0.0010}$ & $0.1194^{+0.0020}_{-0.0020}$ & $0.1193^{+0.0009}_{-0.0009}$ & $0.1193^{+0.0018}_{-0.0019}$ \\
\rowcolor{gray!10}
$100~\theta_{s}$ & $1.042^{+0.0003}_{-0.0003}$ & $1.042^{+0.0006}_{-0.0006}$ & $1.042^{+0.0003}_{-0.0003}$ & $1.042^{+0.0006}_{-0.0005}$ \\
${\rm ln}(10^{10}A_{\rm s })$ & $3.044^{+0.016}_{-0.016}$ & $3.044^{+0.033}_{-0.032}$ & $3.037^{+0.016}_{-0.015}$ & $3.037^{+0.031}_{-0.031}$ \\
\rowcolor{gray!10}
$n_{s}$ & $0.9650^{+0.0041}_{-0.0045}$ & $0.9650^{+0.0079}_{-0.0103}$ & $0.9660^{+0.0039}_{-0.0040}$ & $0.9660^{+0.0076}_{-0.0090}$ \\
$\tau_{ reio}$ & $0.0551^{+0.0076}_{-0.0074}$ & $0.0551^{+0.0159}_{-0.0148}$ & $0.0520^{+0.0074}_{-0.0071}$ & $0.0520^{+0.0150}_{-0.0149}$ \\
\rowcolor{gray!10}
$\log_{10}z_{\rm int}$ & $>5.22$ & $>4.89$ & $>5.49$ & $>5.11$ \\
$\sum m_\nu[{\rm eV}]$ & $<0.09$ & $<0.15$ & $<0.11$ & $<0.16$ \\
\rowcolor{gray!10}
$H_0$ & $67.83^{+0.50}_{-0.52}$ & $67.83^{+0.97}_{-1.06}$ & $67.69^{+0.42}_{-0.42}$ & $67.69^{+0.82}_{-0.85}$ \\
\hline
\end{tabularx}
}
\caption{\it Cosmological and model parameter constraints with mean values and \textbf{68\% and  95\% C.L.} for $n_{\rm int}=4$ model. The limits are shown with \textbf{Planck+BAO} and \textbf{Planck+BAO+FS} datasets.}
\label{tab:nint4}
\end{table}

\begin{table}[h!]
\centering
\renewcommand{\arraystretch}{1.5} 
\resizebox{\linewidth}{!}{ 
\begin{tabularx}{\textwidth}{|>{\centering\arraybackslash}p{2.2cm}|X|X|X|X|}
\hline
\multirow{2}{*}{\textbf{Parameters}} & \multicolumn{2}{c|}{Planck+BAO} & \multicolumn{2}{c|}{Planck+BAO+FS} \\
\cline{2-5}
& \textbf{Mean} $\bm{\pm \; 1 \sigma} $ & \textbf{Mean} $\bm{\pm \; 2 \sigma }$ & \textbf{Mean} $\bm{\pm \; 1 \sigma} $ & \textbf{Mean} $\bm{\pm \; 2 \sigma}$ \\
\hline
\rowcolor{gray!10}
$100~\omega_{\rm b}$ & $2.241^{+0.013}_{-0.014}$ & $2.241^{+0.027}_{-0.026}$ & $2.239^{+0.013}_{-0.012}$ & $2.239^{+0.026}_{-0.024}$ \\
$\omega_{\rm cdm}$ & $0.1195^{+0.0010}_{-0.0010}$ & $0.1195^{+0.0020}_{-0.0020}$ & $0.1194^{+0.0008}_{-0.0009}$ & $0.1194^{+0.0015}_{-0.0018}$ \\
\rowcolor{gray!10}
$100~\theta_{s}$ & $1.042^{+0.0003}_{-0.0003}$ & $1.042^{+0.0005}_{-0.0005}$ & $1.042^{+0.0002}_{-0.0003}$ & $1.042^{+0.0005}_{-0.0005}$ \\
${\rm ln}(10^{10}A_{\rm s })$ & $3.044^{+0.016}_{-0.016}$ & $3.044^{+0.033}_{-0.032}$ & $3.040^{+0.013}_{-0.014}$ & $3.040^{+0.027}_{-0.027}$ \\
\rowcolor{gray!10}
$n_{s}$ & $0.9651^{+0.0042}_{-0.0047}$ & $0.9651^{+0.0080}_{-0.0102}$ & $0.9666^{+0.0036}_{-0.0034}$ & $0.9666^{+0.0067}_{-0.0066}$ \\
$\tau_{ reio}$ & $0.0549^{+0.0079}_{-0.0072}$ & $0.0549^{+0.0161}_{-0.0152}$ & $0.0538^{+0.0068}_{-0.0067}$ & $0.0538^{+0.0129}_{-0.0128}$ \\
\rowcolor{gray!10}
$\log_{10}z_{int}$ & $>5.19$ & $>4.94$ & $>5.60$ & $>5.23$ \\
$\sum m_\nu[{\rm eV}]$ & $<0.08$ & $<0.14$ & $<0.11$ & $<0.17$ \\
\rowcolor{gray!10}
$H_0$ & $67.82^{+0.48}_{-0.51}$ & $67.82^{+0.94}_{-1.02}$ & $67.64^{+0.37}_{-0.40}$ & $67.64^{+0.74}_{-0.80}$ \\
\hline
\end{tabularx}
}
\caption{\it Cosmological and model parameter constraints with mean values and \textbf{68\% and  95\% C.L.} for $n_{\rm int}=5$ model. The limits are shown with \textbf{Planck+BAO} and \textbf{Planck+BAO+FS} datasets.}
\label{tab:nint5}
\end{table}

\begin{figure}[ht!]
    \centering
    \includegraphics[width=\textwidth]{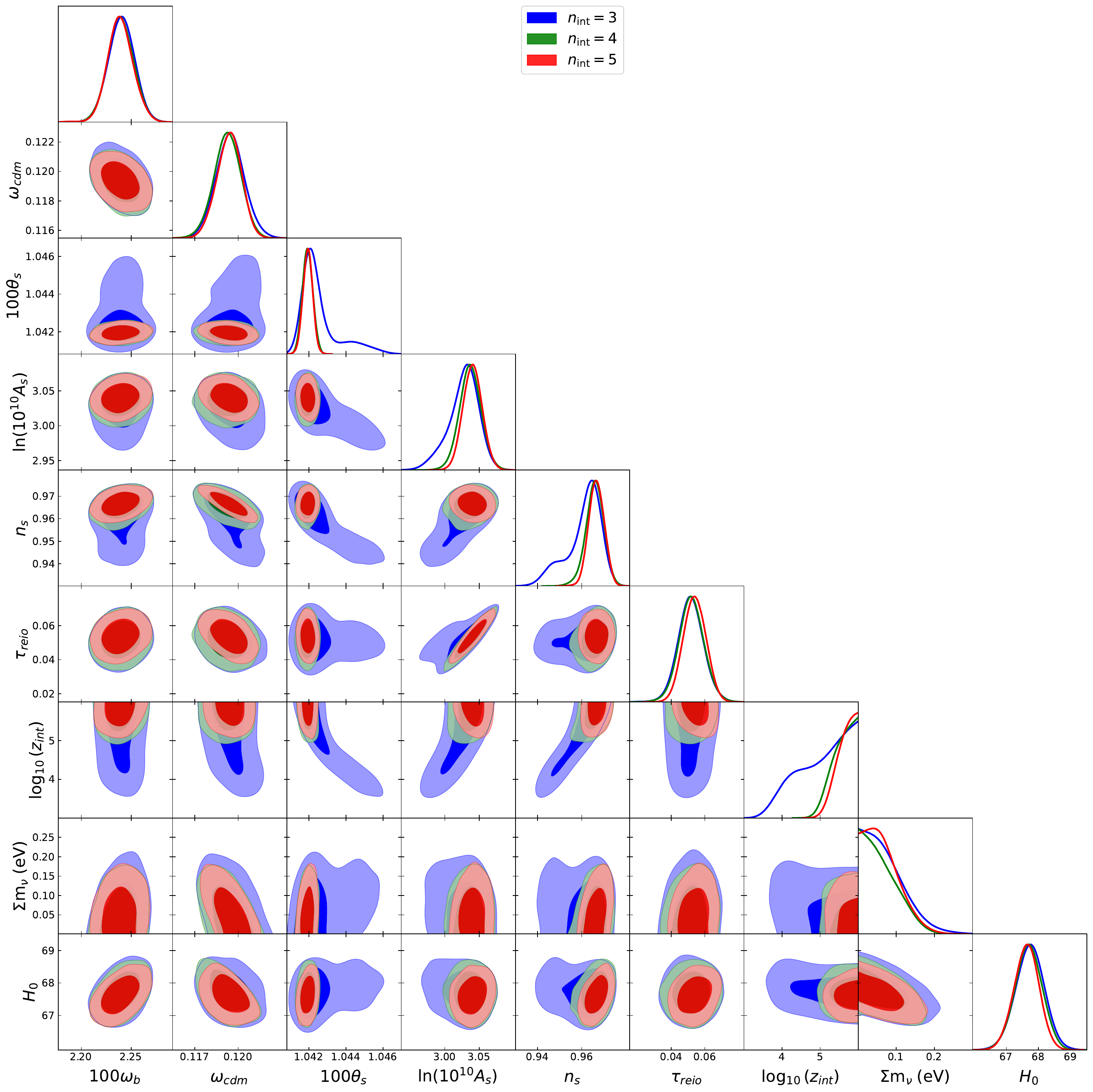}
    \caption{\it Comparative analysis of triangular plots for all the  cosmological and model parameters with mean values and \textbf{68\% and  95\% C.L.} for $n_{\rm int}=3,4$ and $5$ cases (blue, green and red respectively) with \textbf{Planck+BAO+FS} datasets.}
    \label{fig:2d_posterior_all_nint}
\end{figure}

Furthermore, Fig.~\ref{fig:2d_posterior_all_nint} shows a comparison of the full posterior probability distribution with $1\sigma$ and $2\sigma$ uncertainty of all the parameters under consideration, for three different cases of interaction under consideration, using Planck + BAO +FS dataset. The contours in blue, green and red respectively represent the models with $n_{\rm int} =3,\, 4$ and $5$. As mentioned earlier, it is important to note that the $n_{\rm int}=5$ model effectively maps to the moderately interacting (MI) mode studied in \cite{Camarena:2023cku,Camarena:2024zck}\footnote{The interaction strength obtained in our analysis closely lies within the range obtained in \cite{Camarena:2023cku}. That's why we are referring it as moderately interacting mode.}.

It is crucial to investigate the impact on cosmological parameters other than the interaction redshift, $z_{\rm int}$, when incorporating full shape data. As demonstrated in \cite{Ivanov:2019hqk}, the geometrical measurements of BAO provide nearly equivalent information to the broadband shape data in the context of BOSS DR12 dataset. Consequently, we obtain comparable constraints on the cosmological parameters with including shape information from FS data. Important point to notice in all of these cases is that, since FS data is insensitive to the optical depth and sum of neutrino mass, including this dataset on top of BAO loosens the constraint on $\tau$ slightly (within $1\sigma$) and for sum of neutrino mass (within $2\sigma$). Although for $n_{\rm int}=3$ model, the degeneracies between the parameters have already been identified in \cite{Taule:2022jrz}, here our analysis with FS data show identical degeneracies with slight improvements.  The key constraint of interest, the interaction redshift, is provided separately in Eq.~\ref{eq:bounds_fs}. 
The constraints on $z_{\rm int}$ from Planck+BAO+FS dataset at 95\% C.L., are as follows:
\begin{subequations}
\label{eq:bounds_fs}
\begin{align}
   n_{\rm int} &= 3: \,\,\, z_{\rm int} >7.93 \times 10^3 \, , \\
   n_{\rm int} &= 4: \,\,\, z_{\rm int} >1.28 \times 10^5 \, , \\
   n_{\rm int} &= 5: \,\,\, z_{\rm int} >1.7 \times 10^5 \, .
\end{align}
\end{subequations}

Further, it is important to note that the inclusion of FS datasets leads to a relaxation in the bounds on the sum of neutrino masses, consistent with the findings in \cite{Philcox:2020vvt,Philcox:2021kcw}. With Planck +BAO +FS dataset, we obtain the bounds on sum of neutrino mass as $\sum m_\nu< 0.19\,{\rm eV}$ and $\sum m_\nu< 0.16\,{\rm eV}$ for $n_{\rm int}=3$ and $4$ cases respectively. For $n_{\rm int}=5$ case  we obtain $\sum m_\nu< 0.16\,{\rm eV}$ at 95\% C.L. As noted in \cite{Philcox:2020vvt,Philcox:2021kcw}, including FS data with Planck and BAO slightly loosens the bounds on $\sum m_\nu$ as opposed to Planck + BAO analysis for $\Lambda$CDM +$\sum m_\nu$ model. This adjustment slightly alters the onset of neutrino free-streaming across all cases.

For $n_{\rm int}=5$ model, using dimensional analysis, the neutrino interaction rate can be expressed as $\Gamma_\nu \simeq G_{\rm eff}^2 T_\nu^5$, where $\Gamma_\nu = H(z_{\rm int})$. This relationship allows us to infer the neutrino interaction strength based on the Hubble parameter at the interaction redshift. Analysis combining Planck and BAO data yields a 95\% C.L. constraint of $z_{\rm int} > 8.7 \times 10^4$, which corresponds to an upper limit on the interaction strength parameter: $G_{\rm eff} < 3.9 \times 10^{-4} \, {\rm MeV}^{-2}$. Including the BOSS DR12 Full Shape galaxy power spectra further tightens this constraint to $z_{\rm int} > 1.7 \times 10^5$ and $G_{\rm eff} < 1.59 \times 10^{-4} \, {\rm MeV}^{-2}$. Given the insufficiency in the literature of obvious mapping to any specific particle physics model for $n_{\rm int} = 3$ and 4 cases, our analysis for these two cases focuses mostly on determining constraints on the interaction redshift along with standard cosmological parameters. As presented in Tables~\ref{tab:nint3} and \ref{tab:nint4}, we find that $z_{\rm int} > 7.93 \times 10^3$ for $n_{\rm int} = 3$ and $z_{\rm int} > 1.28 \times 10^5$ for $n_{\rm int} = 4$ at a 95\% C.L.

\section{Forecasts on future CMB+LSS missions}
\label{sec:forecast}

Future CMB experiments such as CMB-S4 \cite{CMB-S4:2016ple,abitbol2017cmb}, PICO \cite{Sutin:2018onu,young2018optical} and LiteBIRD \cite{Matsumura:2013aja,suzuki2018litebird}, along with future LSS experiment like Euclid \cite{EuclidTheoryWorkingGroup:2012gxx,Amendola:2016saw} for galaxy redshift surveys, are expected to provide crucial insights into constraining various cosmological parameters. The combination of both future CMB and LSS missions will help in probing the mildly non-linear regime and the dynamics on very small scales with unprecedented sensitivity. Keeping this in mind, we proceed to perform a forecast analysis of the early universe neutrino interaction scenarios as discussed in the present article, in the context of the upcoming data from CMB-S4 and Euclid. We choose CMB-S4 and Euclid as both of these missions are aimed to probe sum of neutrino mass with an unprecedented accuracy. 

For the forecast analysis with Euclid galaxy clustering we compute the galaxy power spectra from the matter power spectra using the simple relation,
\begin{equation}
\label{galaxy_ps_halofit}
P_g(k,\mu,z) =  f_{\text{AP}}(z) \times f_{\text{res}}(k,\mu,z) \times f_{\text{RSD}}(\hat{k},\hat{\mu},z) \times b^2(z) \times P_m(\hat{k},z) \ .
\end{equation}
where $\mu$ is the angle between the wave vector and the observer's line of sight vector and the computation is performed under the flat-sky approximation. The $f_i$ terms in the above equation are described as follows:

\begin{enumerate}[*]
    \item \textit{Alcock-Paczynski Effect:} The term $f_{\text{AP}}(z) = D_A^2 \hat{H}/{\hat{D}_A^2 H}$ accounts for the difference between the true cosmology and fiducial cosmology. Here $D_A$ and $H$ are the angular diameter distance and Hubble parameter respectively. Here wherever there is a `hat', it represents the true cosmology and `A' denotes assumed (fiducial) cosmology.
    \item \textit{Instrument Resolution Effect:} Due to the limited resolution of instruments, small scale perturbations are suppressed. Assuming Gaussian errors for the coordinate parallel ($\sigma_{\shortparallel}$) and perpendicular ($\sigma_{\perp}$) to the line of sight at redshift z, the suppression factor is:
    \begin{equation}
    f_{\text{res}}(k,\mu,z) = \exp\left(-k^2\left[\mu^2\cdot\left(\sigma_{\shortparallel}^2(z)-\sigma_{\perp}^2(z)\right)+\sigma_{\perp}^2(z)\right]\right) \ .
    \end{equation}
    Here, the parallel positional uncertainty is given by $\sigma_{\shortparallel} = \frac{c}{H} \sigma_z $, where $\sigma_z$ is the uncertainty in redshift. The effect of angular resolution uncertainty is neglected, allowing us to set $\sigma_{\perp}$ zero following \cite{audren2013neutrino}.
    
    \item \textit{RSD:} On large scales within the linear regime, an apparent anisotropy is induced in redshift space because of the classical Doppler effect. This is known as the Kaiser formula \cite{Kaiser:1987qv} that is described by the first term here. On top of this large scale effect, peculiar velocities of galaxies further distorts the appearance in redshift space along the line of sight on small scales. This feature is called fingers of God \cite{Jackson:1971sky}. Following \cite{Bull:2014rha}, we can describe this additional suppression with an exponential factor. Including both the large and small scale effects, the RSD correction can be expressed as,  $f_{\text{RSD}}(\hat{k},\hat{\mu},z) = \left( 1+\beta(\hat{k},z) \, {\hat{\mu}}^2 \right)^2 e^{-{\hat{k}}^2{\hat{\mu}}^2\sigma_{\text{NL}}^2}$.   
\end{enumerate}

Given that let us now focus on the noise spectra for Euclid.
We divide the surveys into bins of width $\Delta z = 0.1$ with mean redshift $\bar{z}$. Correlation functions are defined inside the bin's data and are approximated to probe the power spectrum at a fixed redshift $\bar{z}$. The volume of one redshift bin can be computed via
\begin{equation}
V_r(\bar{z}) = 4\pi f_{\text{sky}}\cdot\int_{\Delta r(\bar{z})}r^2 { dr} = \frac{4\pi}{3} f_{\text{sky}} \cdot\left[r^3\left(\bar{z}+\frac{\Delta z}{2}\right)-r^3\left(\bar{z}-\frac{\Delta z}{2}\right)\right] \ ,
\label{eq:Vr}
\end{equation}
where $f_{\text{sky}}$ is the fraction of the sky covered by the survey.
The distribution of galaxies is discrete, rather than continuous like the density field $\delta_g$.
Thus we have to take into account the experimental shot noise in each redshift bin. 
\begin{equation}
P_{N}(\bar{z}) = \frac{1}{\bar{n}(\bar{z})} = \frac{V_r(\bar{z})}{N(\bar{z})} \ ,
\end{equation}
where $N(\bar{z})$ is the number of galaxies in the bin, $V_r(\bar{z})$ the volume of the bin and $\bar{n}(\bar{z})$ the galaxy number density. 

Taking this shot noise into account, the observed power spectra in each bin is actually the sum of the theoretical galaxy power and the shot noise.
\begin{equation}
P_{\mathrm{obs}}(k,\mu,\Bar{z}) = P_g(k,\mu,\Bar{z}) + P_{N}(\Bar{z})
\end{equation}

\subsection{Euclid specifications}

Euclid satellite promises to measure galaxy clustering and weak gravitational lensing with better than 1\% accuracy enabling more precise measurement of cosmological parameters \cite{EuclidTheoryWorkingGroup:2012gxx,Amendola:2016saw}. This will significantly enhance our understanding of dark matter distribution, dark energy as well as their interplay with other cosmic species. By performing a spectroscopic measurement, Euclid aims to gather data from approximately $10^{7}$ galaxies across a redshift range of $0.7-2.0$. It should be noted here that in the present analysis, we have used only the spectroscopic measurements. The photometric probes of Euclid mission are not considered here.
The error in spectroscopic measurements for this survey can be described by $\sigma_{\rm z} = 0.001(1+z)$ from which we calculate the parallel positional uncertainty, while angular resolution errors have been neglected \cite{EuclidTheoryWorkingGroup:2012gxx,audren2013neutrino}. Detailed specifications of Euclid are listed in Table. \ref{EuclidSpec}. Euclid will detect galaxies over a sky fraction $f_{\rm sky}=0.3636$, within redshift bins of width $\Delta z$ centered around $\Bar{z}$, as described by,

\begin{equation}
 N(\Bar{z}) = 41253 f_{\rm sky} ~\rm deg^{2} \int_{\Bar{z}-\rm \Delta z/2}^{\Bar{z}+\rm \Delta z/2} \frac{dN(z)/dz}{1~\rm deg^{2}} dz.  
\end{equation}
Once obtained the number of galaxies within a redshift bin from the above equation the shot noise is calculated using the relation $\frac{1}{\bar{n}(\bar{z})} = \frac{V_r(\bar{z})}{N(\bar{z})}$, where $V_r(\bar{z}$ is calculated from the Eqn.~\ref{eq:Vr}. Since the galaxy bias is redshift dependent, following \cite{audren2013conservative}, we modeled the bias by the simple relation $b(z)=\sqrt{1+z}$. To account the uncertainty in assuming the simple relation two additional nuisance parameters $\beta_{\rm 0}^{\rm Euclid}$ and $ \beta_{\rm 1}^{\rm Euclid}$, have been introduced in modeling the galaxy bias factor detected by Euclid \cite{audren2013neutrino}, expressed as,

\begin{equation}
    b_{z} = \beta_{\rm 0}^{\rm Euclid}(1+z)^{0.5 \beta_1^{\rm Euclid}}.
\end{equation}
As a prior we have chosen Gaussian priors with $\sigma=2.5\%$ for these $\beta$ parameters.

\begin{table}
\begin{center}
\caption{Euclid specifications.}
\label{EuclidSpec}
\begin{tabular}{c|ccccc} 
 \toprule 
parameter & $z_{\text{min}}$ & $z_{\text{max}}$ & $f_{\text{sky}}$ & $\sigma_{z}$ &$\sigma_{\theta}$ [$''$]\\
\midrule 
Euclid & 0.7 & 2.0 & 0.3636 & $0.001(1+z)$ & 0 \\
\bottomrule 
\end{tabular}
\end{center}
\end{table}

\subsection{CMB-S4 specifications}
In our forecast analysis, we also include  CMB-S4 as the future CMB mission.
CMB-S4 is the first ground based stage-IV CMB project with the primary goal to search for inflationary B modes. Along with that it will also be able to measure the sum of neutrino mass with a target threshold of $2\sigma$ and $3\sigma$ detection to $0.03{\rm eV}$ and $0.02{\rm eV}$ respectively \cite{CMB-S4:2016ple,abitbol2017cmb}. In the CMB maps, multipole moments receive contributions primarily from the CMB signal $s_{\ell m}$ and the experimental noise $n_{\ell m}$, which can be written as,

\begin{equation}
    a_{\ell m}^P = s_{\ell m}^P + n_{\rm \ell m}^P
\end{equation}

Here $P$ stands for temperature and E and B polarization modes respectively. The noise spectrum for CMB-S4 can be modeled as following \cite{Perotto:2006rj,Brinckmann:2018owf},

\begin{equation}
N_\ell^{PP'} \equiv \langle n_{\ell m}^{P*} n_{\ell m}^{P'} \rangle =
\delta_{P P'} \
\theta_{\rm{FWHM}}^2 \
\sigma_P^2 \ \exp\left[\ell(\ell+1) \ \frac{\theta_{\rm{FWHM}}^2}{8 \ln 2}\right],
\end{equation}
where $\theta_{\rm FWHM}$ and $\sigma_P$ represent the full width at half maximum of the Gaussian beam and root mean square of the instrumental noise. 
CMB-S4 is designed to probe at a target frequency 150 GHz with a beam width 3.0 arcmin and temperature and polarization sensitivity 1.0 and 1.41 $\mu$K arcmin respectively \cite{Perotto:2006rj,Brinckmann:2018owf}.

\begin{figure}[ht!]
    \centering
    \includegraphics[width= 0.65 \textwidth]{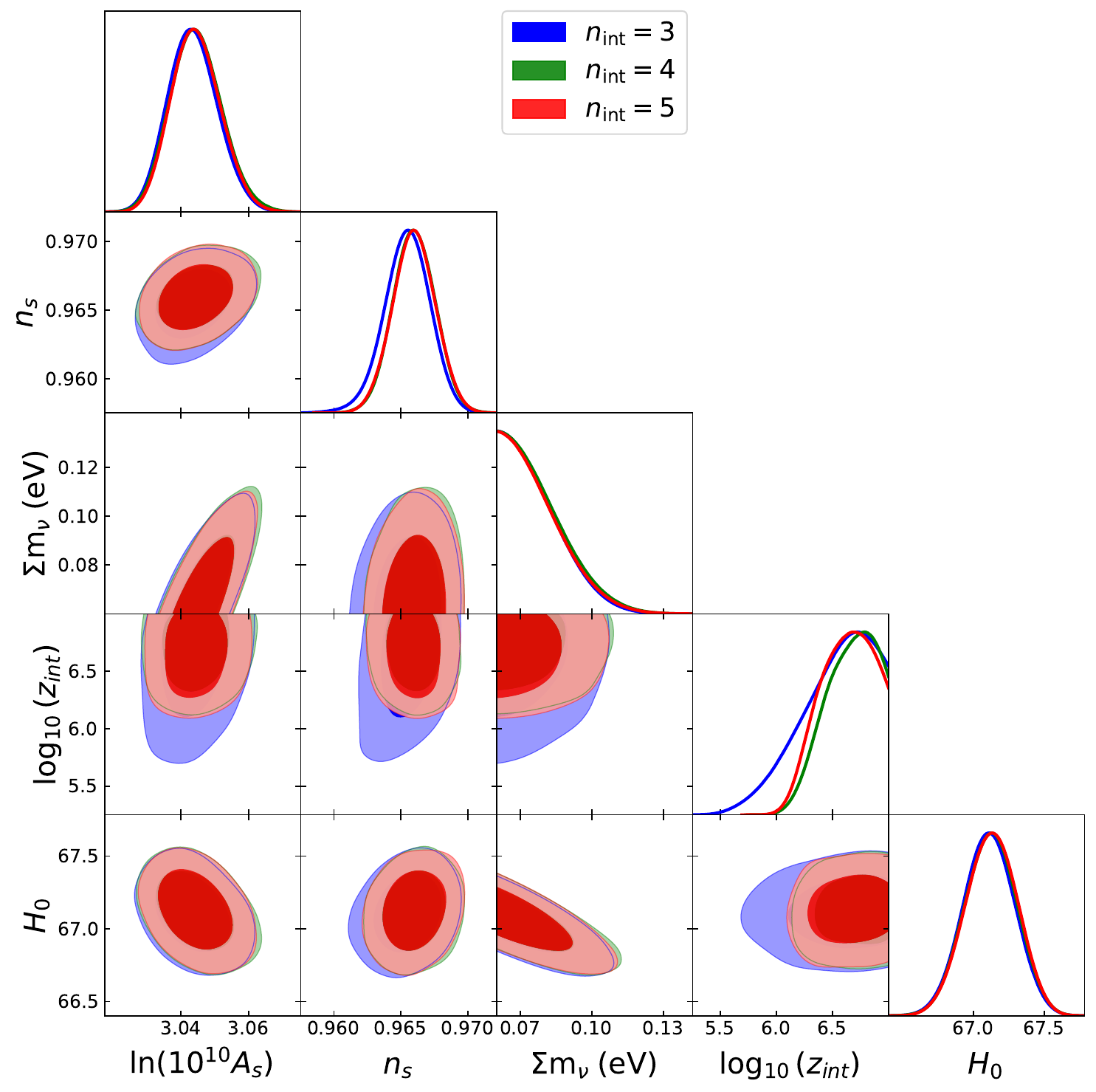}
    \caption{\it Triangular plots representing the 2D and 1D marginalized posterior posterior distributions for $A_s, n_s,H_0,\sum m_\nu$ and $z_{\rm int}$ from \textbf{Planck+CMB-S4+Euclid} forecast analysis for $n_{\rm int}=3,4$ and $5$ models (blue, green and red respectively). Detailed triangular plots for each individual case have been shown in Appendix~\ref{appendix:b}.}
    \label{fig:euclid_analysis}
\end{figure}

\subsection{Covariance matrix and MCMC analysis}
The focus of our study is to include the non-linear corrections to the matter power spectrum in our forecast analysis. There has been a handful of studies to search for possible constraints on the sum of neutrino mass based on the sensitivity of Euclid, incorporating these non-linear corrections \cite{Hamann:2012fe,audren2013neutrino,Chudaykin:2019ock,Euclid:2024imf}. Our analysis may be considered as a complimentary to that, where possible neutrino interactions have also been taken into account in a model-independent way.  Unlike the approach in \cite{Chudaykin:2019ock}, which employs the EFT of LSS for Euclid specifications, we apply the standard \textit{Halofit} \cite{Smith:2002dz,bird2012massive,Takahashi:2012em} corrections to the matter power spectrum and the corresponding Euclid sensitivity model as previously described. This choice is made primarily to minimize the number of nuisance parameters in the analysis. We caution the reader to refrain from making any comparison between the investigations carried out in Sec.~\ref{sec:results_analysis}, where we used EFT of LSS framework for the non-linear modeling and the analysis carried out in this section where we make use the \textit{Halofit} for modeling non-linear regime.
Within the \texttt{CLASS-PT} we use the non-linear flag set to\textit{Halofit} and restrict the $k_{\rm max}$ to a conservative limit to $0.2\, h/{\rm Mpc}$ to minimize error (since  the error is found to increase with further increase of $k_{\rm max}$ value \cite{Sprenger:2018tdb}.)

\begin{table}[h!]
\centering
\renewcommand{\arraystretch}{1.5}
\begin{tabularx}{\textwidth}{|>{\centering\arraybackslash}p{2.5cm}|>{\centering\arraybackslash}X|>{\centering\arraybackslash}X|}
\hline
\textbf{Parameters} & \multicolumn{2}{c|}{\textbf{Planck+CMB-S4+Euclid ($n_{\rm int}=3$)}} \\
\hline
& \textbf{Mean} $\bm{\pm \; 1\sigma}$ & \textbf{Mean} $\bm{\pm \; 2\sigma}$ \\
\hline
$100~\omega_{\rm b}$ & $2.236^{+0.003}_{-0.003}$ & $2.236^{+0.006}_{-0.006}$ \\
$\omega_{\rm cdm}$ & $0.1202^{+0.0003}_{-0.0003}$ & $0.1202^{+0.0006}_{-0.0006}$ \\
$100~\theta_{s}$ & $1.041^{+0.0001}_{-0.0001}$ & $1.041^{+0.0002}_{-0.0002}$ \\
${\rm ln}(10^{10}A_{\rm s})$ & $3.043^{+0.007}_{-0.007}$ & $3.043^{+0.015}_{-0.013}$ \\
$n_{s}$ & $0.9655^{+0.0016}_{-0.0017}$ & $0.9655^{+0.0032}_{-0.0036}$ \\
$\tau_{ reio}$ & $0.0539^{+0.0041}_{-0.0038}$ & $0.0539^{+0.0082}_{-0.0072}$ \\
$\log_{10}z_{\rm int}$ & $>6.18$ & $>5.80$ \\
$\sum m_\nu[{\rm eV}]$ & $<0.079$ & $<0.099$ \\
$H_0$ & $67.11^{+0.18}_{-0.18}$ & $67.11^{+0.35}_{-0.35}$ \\
\hline
\end{tabularx}
\caption{\it Cosmological and model parameter constraints with mean values and \textbf{68\% and 95\% C.L.} for $n_{\rm int}=3$ model with \textbf{Planck+CMB-S4+Euclid} forecast analysis.}
\label{tab:euclid_forecast_1}
\end{table}

\begin{table}[h!]
\centering
\renewcommand{\arraystretch}{1.5} 
\resizebox{\linewidth}{!}{ 
\begin{tabularx}{\textwidth}{|>{\centering\arraybackslash}p{2.2cm}|X|X|X|X|}
\hline
\multirow{2}{*}{\textbf{Parameters}} & \multicolumn{2}{c|}{Planck+CMB-S4+Euclid ($n_{\rm int}=4$)} & \multicolumn{2}{c|}{Planck+CMB-S4+Euclid ($n_{\rm int}=5$)} \\
\cline{2-5}
& \textbf{Mean} $\bm{\pm \; 1 \sigma} $ & \textbf{Mean} $\bm{\pm \; 2 \sigma }$ & \textbf{Mean} $\bm{\pm \; 1 \sigma} $ & \textbf{Mean} $\bm{\pm \; 2 \sigma}$ \\
\hline
\rowcolor{gray!10}
$100~\omega_{\rm b}$ & $2.237^{+0.003}_{-0.003}$ & $2.237^{+0.006}_{-0.006}$ & $2.237^{+0.003}_{-0.003}$ & $2.237^{+0.006}_{-0.006}$ \\
$\omega_{\rm cdm}$ &$0.1202^{+0.0003}_{-0.0003}$ & $0.1202^{+0.0006}_{-0.0006}$  & $0.1201^{+0.0003}_{-0.0003}$ & $0.1201^{+0.0006}_{-0.0006}$  \\
\rowcolor{gray!10}
$100~\theta_{s}$ & $1.041^{+0.000085 }_{-0.000085 }$ & $1.041^{+0.000163 }_{-0.000163 }$ & $1.041^{+0.00008 }_{-0.00008}$ & $1.041^{+0.00016 }_{-0.00016}$ \\
${\rm ln}(10^{10}A_{\rm s })$ & $3.044^{+0.008}_{-0.007}$ & $3.044^{+0.016}_{-0.013}$ & $3.044^{+0.007}_{-0.007}$ & $3.044^{+0.015}_{-0.013}$ \\
\rowcolor{gray!10}
$n_{s}$ & $0.9660^{+0.0031}_{-0.0030}$ & $0.9666^{+0.0067}_{-0.0066}$ & $0.9660^{+0.0016}_{-0.0015}$ & $0.9660^{+0.0031}_{-0.0031}$ \\
$\tau_{ reio}$ & $0.0540^{+0.0042}_{-0.0040}$ & $0.0540^{+0.0087}_{-0.0075}$  & $0.05401^{+0.0042}_{-0.0039}$ & $0.0540^{+0.0083}_{-0.0074}$ \\
\rowcolor{gray!10}
$\log_{10}z_{\rm int}$ & $>6.45$ & $>6.20$ & $>6.41$ & $>6.20$ \\
$\sum m_\nu[{\rm eV}]$ &$<0.082$ & $<0.10$ & $<0.08$ & $<0.10$ \\
\rowcolor{gray!10}
$H_0$ & $67.12^{+0.18}_{-0.19}$ & $67.12^{+0.35}_{-0.36}$ & $67.13^{+0.18}_{-0.18}$ & $67.13^{+0.34}_{-0.36}$ \\
\hline
\end{tabularx}
}
\caption{\it Cosmological and model parameter constraints with mean values and \textbf{68\% and  95\% C.L.} for $n_{\rm int}=4$ and 5 models with \textbf{Planck+CMB-S4+Euclid} forecast analysis.}
\label{tab:euclid_forecast_2}
\end{table}

Let us now turn to briefly discuss about the Fisher matrix method and Markov Chain Monte Carlo (MCMC) analysis for the future experiments taken into our analysis. Fisher matrix is an efficient tool to constrain cosmological parameters as well as model parameters with improved accuracy for an upcoming experiment \cite{verde2010statistical}. The Fisher matrix is essentially defined as the second derivative of the log likelihood with respect to the parameters of our interest, evaluated at their best-fit fiducial values.

\begin{equation}\label{Fij}
    F_{ ij}=-\left\langle \dfrac{\partial^{2}\ln\mathcal{L}}{\partial q_{i} \partial q_{j}} \right\rangle = - \dfrac{\partial^{2}\ln\mathcal{L}}{\partial q_{i} \partial q_{j}}~\Bigg|_{q_{0}}.
    \end{equation}

It basically approximates the logarithm of the likelihood function as a multivariate Gaussian function of the parameters $q$ at their fiducial values $q_{0}$. The inverse of the Fisher matrix,  ${\rm Cov}(q_{i},q_{j}) \geq [F^{-1}]_{ij} $ is the covariance matrix. The diagonal elements of this covariance matrix provides the error on the corresponding parameters of our interest. 
   
\begin{equation}
 \sigma(\alpha_{i})=\sqrt{[F^{-1}]_{ii} }   
\end{equation}

Based on the above-mentioned instrumental specifications and possible sources of error, we 
first compute the covariance matrix with the fiducial values of the cosmological parameters set as mentioned in the footnote. We then perform Bayesian MCMC analysis with the updated covariance matrix for our $6+2$ parameters of out model for CMB-S4 and Euclid surveys.
we  generate the CMB temperature and polarization spectra data considering the fake Planck Gaussian likelihood with $f_{\rm sky} =0.57$ for $2<\ell<50$ and CMB-S4 for $f_{\rm sky} =0.4$ for $51<\ell<3000$. For the power spectrum data generation we adopt the conservative approach as in \cite{Sprenger:2018tdb} considering redshift dependent non-linear cut-off $k_{\rm NL}(z)=k_{\rm NL}(0)(1+z)^{2/(2+n_{s})}$ modeling, where $k_{\rm NL(0)}$ is the non-linear cut-off scale today and $n_s$ is the scalar spectral index. In our analysis, we consider $k_{\rm NL}(0)$ to be $0.2 \, h/{\rm Mpc}$ for Euclid.
The relevant parameters are the standard 6 parameters \{$\omega_b, \, \omega_{\rm cdm},\, 100 \theta_{s},$ $ \ln(10^{10}A_s),\, n_{s},\, \tau_{\rm reio} \}$ along with the interaction redshift, $z_{\rm int}$ and sum of neutrino mass, $\sum m_\nu$\footnote{The fiducial parameter values are taken to be: $\omega_b=0.022377,\, \omega_{\rm cdm}=0.1201, \, 100 \theta_{s}=1.0411,\,\ln(10^{10}A_s)=3.0447,\, n_{s}=0.9659,\,\tau_{ reio}=0.543$ and $\sum m_{\nu}=0.06 \;{\rm eV}$}. 

Let us now briefly discuss the major results of our forecast analysis as presented 
in Fig.~\ref{fig:euclid_analysis} (while the constraints on whole set of parameters are presented in Fig.~\ref{fig:forecast_euclid} in Appendix~\ref{appendix:b}) as well as  
in Tables~\ref{tab:euclid_forecast_1}, \ref{tab:euclid_forecast_2}. It has been investigated earlier in \cite{Taule:2022jrz}  that
CMB-S4 in combination with Planck Baseline, have the potential to constrain the onset of neutrino free-streaming to $z_{\rm int}>2.4 \times 10^5$ for $n_{\rm int} =3$ and $z_{\rm int}>2.8 \times 10^5$ for $n_{\rm int} =5$ at 95\% C.L. CMB-S4 will also be able to break the degeneracy of $z_{\rm int}$ with $A_s,\,n_s$ and $H_0$ for $n_{\rm int}=3$ case. With the inclusion of Euclid, our analysis goes over the previous literature. Since increasing  the value of $z_{\rm int}$ essentially affects the small scales in the matter power spectra, we obtain a tighter constraint for all the three cases with the  combination of CMB-S4 and Euclid. For $n_{\rm int}=5$ case, our analysis implies earlier decoupling obtaining, $z_{\rm int}>1.78 \times 10^6$ which constrains $G_{\rm eff}<4.3 \times 10^{-6}\, {\rm MeV}^{-2}$. Our analysis thus infers that combining Euclid  with Planck Baseline and CMB-S4 will be able to constrain the neutrino interactions up to redshift,
\begin{subequations}
\label{eq:bounds_forecast}
\begin{align}
   n_{\rm int} &= 3 \,\,\Longrightarrow  \,\,\, z_{\rm int} >6.31 \times 10^5 \,\,\,,\\
   n_{\rm int} &= 4\,\,\,\,\Longrightarrow  \,\,\, z_{\rm int} >1.78 \times 10^6 \,\,,\\
    n_{\rm int} &= 5\,\,\,\,\Longrightarrow  \,\,\, z_{\rm int} >1.78 \times 10^6 \,\,.
\end{align}
\end{subequations}

Additionally, 95\% C.L. bounds for Planck Baseline + CMB-S4 + Euclid improves over the previous forecasts with CMB-S4 alone. Our investigation suggests a joint analysis of CMB-S4 with Euclid would be able to constrain the onset of neutrino free-streaming up to redshift $z \sim 10^6$, implying earlier decoupling even in $\Gamma_\nu \propto T_\nu^3$ model. Forecast analysis with Euclid and CMB-S4 (considering all interaction models considered in this work) will be able to put constraints $\sum m_\nu<0.10\, {\rm eV} $ (at 95 \% C.L.).
From Table~\ref{tab:euclid_forecast_1}, \ref{tab:euclid_forecast_2} we see that $1\sigma$ uncertainties of all cosmological parameters in our neutrino interaction models, with the inclusion of Euclid Galaxy Clustering, remain close to the values predicted as in $\Lambda$CDM + $\sum m_\nu$ model forecast analysis in \cite{Sprenger:2018tdb}. It is important to highlight that our analysis examines different combinations of future experiments compared to \cite{Sprenger:2018tdb}, and therefore, a direct one-to-one comparison is not appropriate. Thus, a joint analysis of Planck, CMB-S4 and Euclid will be able to probe neutrino interactions up to redshift $z \sim 10^6$ and can also put tighter bound on the sum of neutrino mass.

\section{Summary}
\label{sec:summary}

Cosmology has entered a precision era, enabling us to probe particle physics interactions throughout the universe with unprecedented accuracy. In conjunction with CMB observations, Large Scale Structure experiments provide insights into various particle physics models, including neutrino interactions in the early universe. Over the past decade, several studies have suggested that neutrino interactions in the early universe might have been delayed due to yet-to-be-discovered models of neutrino self-interactions, which subsequently affect the evolution of gravitational potentials, leaving detectable imprints on CMB anisotropy and matter power spectra in both linear and mildly non-linear regimes. In the present analysis, we investigated whether LSS data is sensitive to these changes. To do this, we made use of a fairly generic parameterization of  neutrino interaction rates, focusing on the interaction in the early universe, and searched for possible constraints on the (6+2) model parameters using the combined dataset from Planck TT, TE, EE + BAO, along with the full shape (FS) galaxy power spectra data.
 
Analyses using Planck and BAO dataset have placed constraints on the interaction redshifts for neutrino interactions in the early universe, finding $z_{\rm int}>6 \times 10^3$, $7.8 \times 10^4$, and $8.4 \times 10^4$ for models with $n_{\rm int} = 3$, 4, and 5, respectively, at 95\% C.L., consistent with the earlier studies \cite{Taule:2022jrz}. 
Further, since these interactions impact the matter power spectra in the mildly non-linear regime, which is probed by the galaxy full shape power spectra, we included the BOSS DR12 full shape spectra data in our analysis. The FS data was found to have tightened the constraints on the interaction redshifts to $z_{\rm int}>7.93 \times 10^3$, $ 1.28 \times 10^5$, and $1.7 \times 10^5$ for $n_{\rm int} = 3$, 4, and 5 models, respectively, at 95\% C.L. While the inclusion of FS data slightly reduces the degeneracies of $z_{\rm int}$ with cosmological parameters for the $n_{\rm int} =3$ case, it relaxes the bounds on the sum of neutrino mass. We obtained $\sum m_\nu<0.19\,{\rm eV}$ for the $n_{\rm int}=3$ model including the FS data at 95\% C.L. whereas, Planck + BAO data provides a tighter constraint $\sum m_\nu<0.17\,{\rm eV}$. Similar trends in the constraints on sum of neutrino mass persist for $n_{\rm int}=4$ and 5 models.
Furthermore, recent findings suggest that the moderately interacting (MI) mode of neutrino self-interactions mediated by heavy scalars in the early universe (which corresponds to $n_{\rm int} = 5$ in our case) shows a lack of concordance when considering both Planck and LSS data. Our study with galaxy power spectra reveals that an even earlier onset of free-streaming is permitted for the moderately interacting (MI) mode in self-interacting neutrino model. 
 
Having investigated the effects of the present LSS  data, we then moved on to examine the sensitivity of future LSS data to the onset of neutrino free-streaming. Using CMB alone, previous studies found that the upcoming CMB-S4 experiment has the potential to constrain the lower bound of the free-streaming redshift $z_{\rm int}$ to $3 \times 10^5$. Our findings suggest that the Euclid galaxy clustering survey (covering the redshift range $0.7 < z < 2.0$), when combined with Planck and CMB-S4 data, would be able to constrain the interaction strength up to $z_{\rm int} \sim 10^6$. It will further lower the uncertainty on $\sum m_\nu$, leading to $\sigma(\sum m_\nu) \approx 0.02\, {\rm eV}$ with $\sum m_\nu< 0.10\, {\rm eV}$ at 68\% C.L. for almost all the cases. Additionally, the joint forecast study with Planck+CMB-S4+Euclid would help break the parameter degeneracy for the $n_{\rm int} = 3$ model, which persists even in the present Planck+BAO+FS dataset analysis.

In a nutshell, present and future galaxy surveys, combined with CMB missions, play a significant role in shedding light on possible neutrino interaction in the early universe, the sum of the neutrino mass as well as major cosmological parameters. This in turn helps us take a step forward to improve our understanding of the universe and its interplay with this essential particle physics entity  as well as the theories encompassing them. 

The results presented in this article  point to several areas that warrant further investigation. Firstly, since our analysis was conducted using the BOSS DR12 FS datasets, it would be intriguing to explore its extension to the BOSS DR16 full-shape galaxy power spectra, as and when that is made publicly available. A detailed forecast analysis involving both spectroscopic and photometric galaxy clustering measurements, somewhat in the  line of \cite{Baumann:2017gkg} will be an interesting way forward.
Additionally, the study of neutrino interactions requires a robust consideration of all cosmological parameters, particularly the sum of neutrino mass, which remains poorly constrained in the context of the EFT of LSS analysis. While current DESI data releases and the future Euclid survey are expected to provide more precise measurements of the sum of neutrino mass, they have not yet been analyzed within the framework of the EFT of LSS. This lies beyond the scope of our current analysis. A comprehensive study of these surveys, in conjunction with CMB anisotropy datasets, will be crucial for unraveling the complexities of neutrino interactions.

\section*{Acknowledgements}
We would like to thank Petter Taule and David Camarena for useful discussions. SP1  thanks Debarun Paul, Arko Bhaumik, Rahul Shah, Pathikrith Banerjee, Antara Dey and Purba Mukherjee for discussions at various stages of the project. SP1  also  thanks Bithika Halder for constant support and inspiration throughout the project, and CSIR for financial support through Senior Research Fellowship (File no. 09/093(0195)/2020-EMR-I). RS acknowledges support from DST Inspire Faculty fellowship Grant no. IFA19-PH231 at ISI Kolkata and the NFSG Research grant from BITS Pilani Hyderabad. SP2 thanks the Department of Science and Technology, Govt. of India for partial support through Grant No. NMICPS/006/MD/2020-21. 
We gratefully acknowledge the use of the publicly available codes \href{https://github.com/lesgourg/class_public}{\texttt{CLASS}}, \href{https://github.com/Michalychforever/CLASS-PT}{\texttt{CLASS-PT}} and \href{https://github.com/brinckmann/montepython_public}{\texttt{MontePython}} for parameter estimation and \href{https://github.com/cmbant/getdist}{\texttt{Getdist}} for plotting.
We also acknowledge the use of computational facilities of Technology Innovation Hub at ISI Kolkata, along with High Performance Computing facility Pegasus at IUCAA, Pune, India. We thank the anonymous referee for useful comments and suggestions that helped us significantly improve our paper.

\newpage
\appendix
\section{Posterior distribution for all parameters: full shape galaxy spectra}
\label{appendix:a}

Here we show the full posterior probability distribution of all the cosmological and model parameters for each cases (\textit{i.e.} $n_{\rm int}=3,\, 4$ and $5$) with Planck+BAO and Planck+BAO+FS datasets, discussed in detail in Sec.~\ref{sec:results_analysis}.

\begin{figure}[hbt!]
    \centering
    \includegraphics[width=\textwidth]{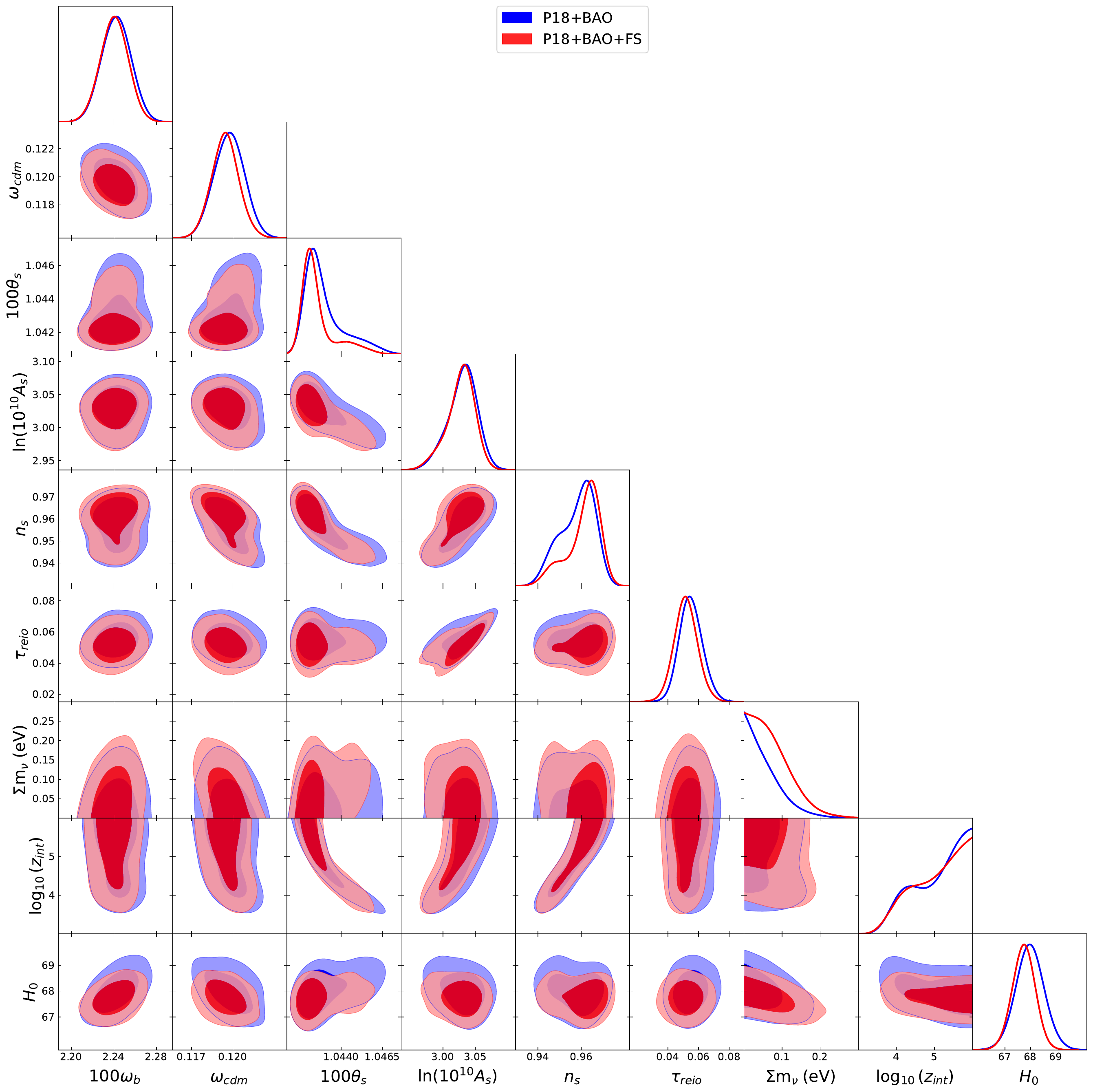}
    \caption{\it \textbf{68\% and 95\% C.L.} marginalized posterior distribution of all cosmological and model parameters with $n_{\rm int}=3$ for \textbf{Planck+BAO} dataset (blue) and \textbf{Planck+BAO+FS} dataset (red).}
    \label{2d_posterior_nint3}
\end{figure}
\FloatBarrier

\begin{figure}[hbt!]
    \centering
    \includegraphics[width=\textwidth]{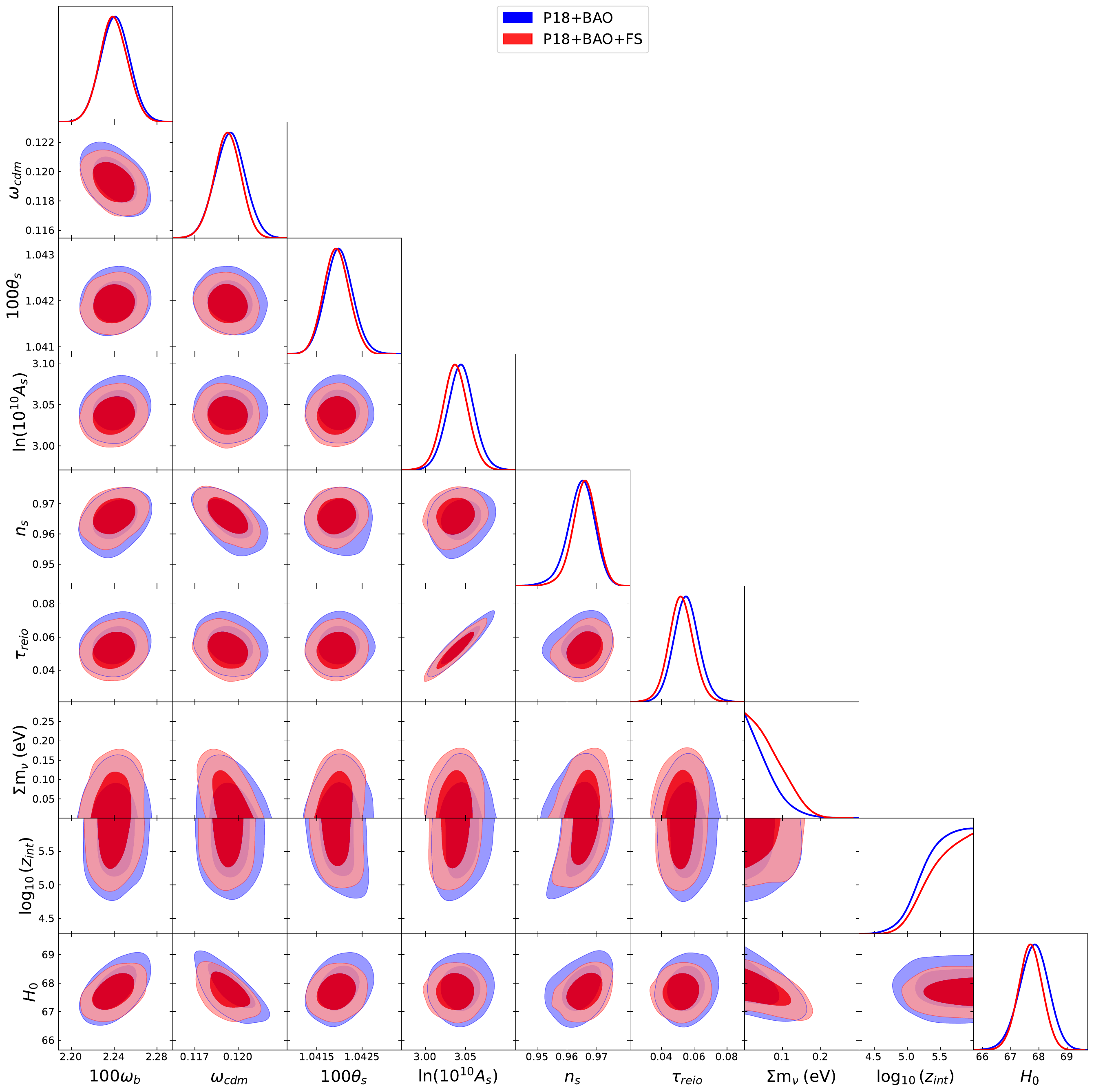}
    \caption{\it \textbf{68\% and 95\% C.L.} marginalized posterior distribution of all cosmological and model parameters with $n_{\rm int}=4$ for \textbf{Planck+BAO} dataset (blue) and \textbf{Planck+BAO+FS} dataset (red).}
    \label{2d_posterior_nint4}
\end{figure}
\FloatBarrier

\begin{figure}[hbt!]
    \centering
    \includegraphics[width=\textwidth]{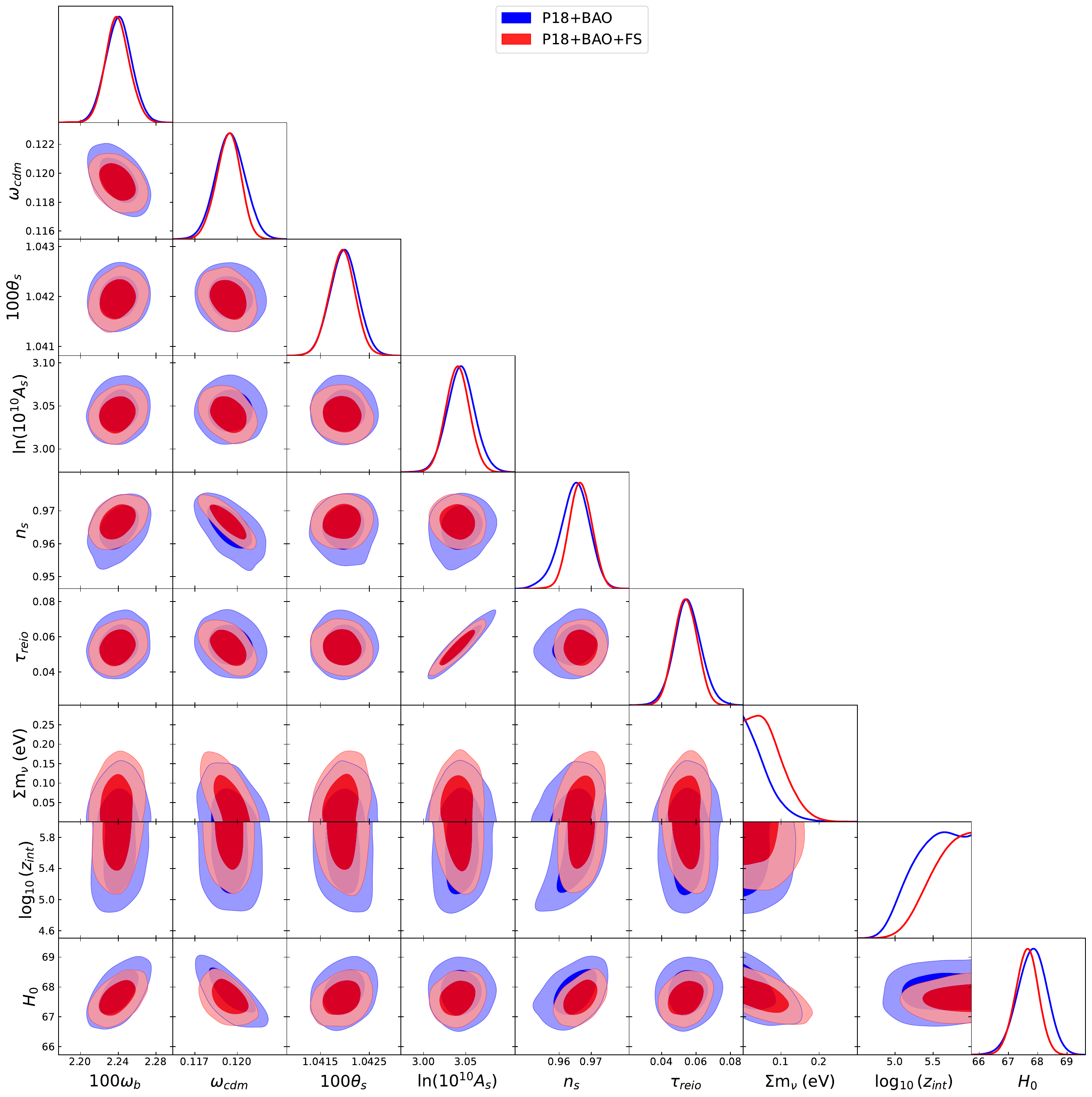}
    \caption{\it \textbf{68\% and 95\% C.L.} marginalized posterior distribution of all cosmological and model parameters with $n_{\rm int}=5$ for \textbf{Planck+BAO} dataset (blue) and \textbf{Planck+BAO+FS} dataset (red).}
    \label{2d_posterior_nint5}
\end{figure}
\FloatBarrier

\section{Posterior distribution for all parameters: future CMB+LSS missions}
\label{appendix:b}

We present here the full posterior probability distribution of all the cosmological parameters for all the models with $n_{\rm int}=3,\, 4$ and $5$ with the forecast analysis as detailed in Sec~\ref{sec:forecast} for Planck+CMB-S4+Euclid.

\begin{figure}[hbt!]
    \centering
    \includegraphics[width=\textwidth]{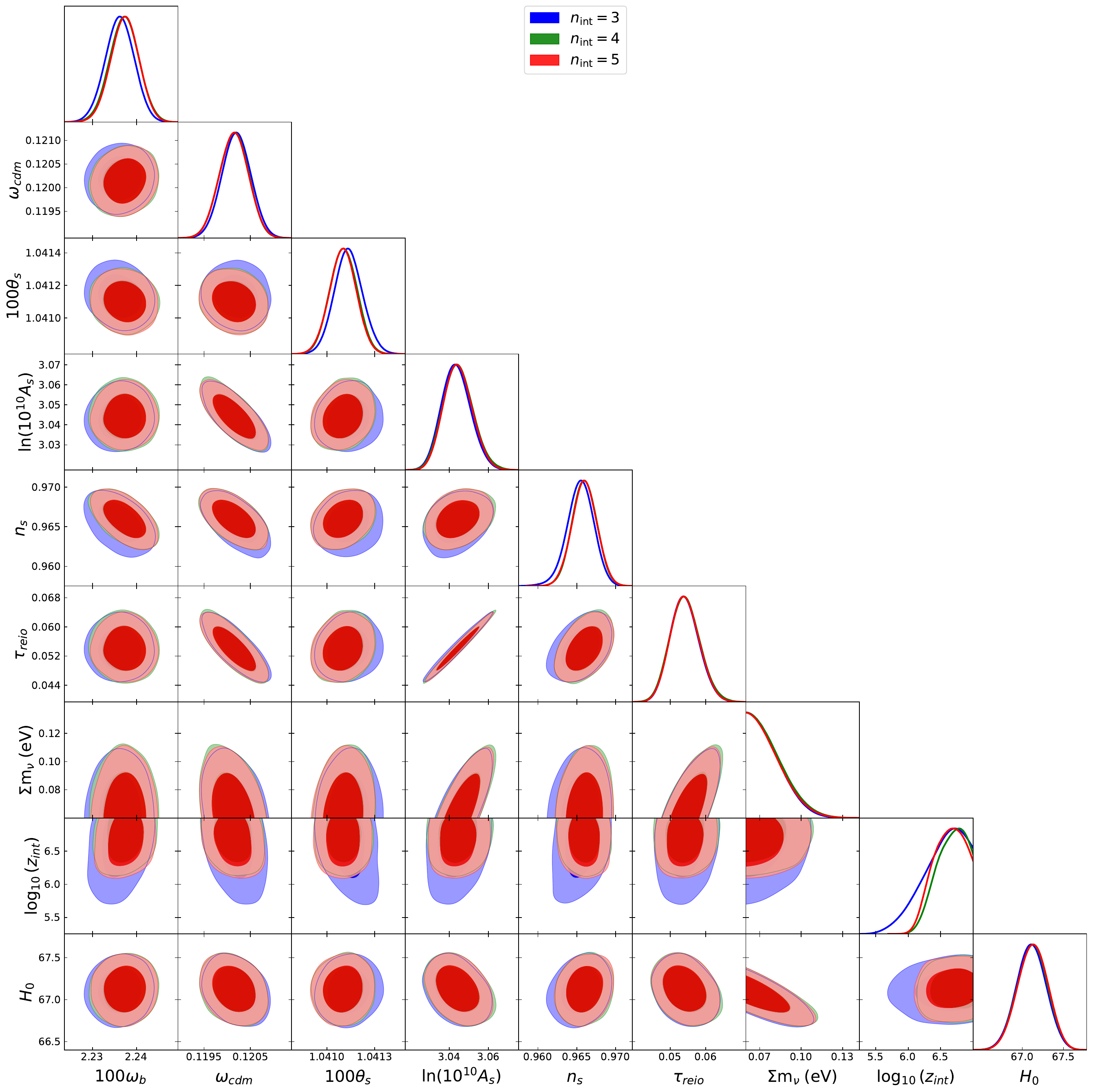}
    \caption{\it \textbf{68\% and 95\% C.L.} posterior probability distribution of all cosmological and model parameters with \textbf{Planck+CMB-S4+Euclid} forecast for $n_{\rm int}=3,4$ and $5$ cases respectively (blue, green and red).}
    \label{fig:forecast_euclid}
\end{figure}
\FloatBarrier

\bibliographystyle{JHEP.bst}
\bibliography{biblio.bib}

\end{document}